\documentclass{mn2e}
\usepackage[utf8x]{inputenc}
\usepackage{float}
\usepackage{graphicx}
\usepackage{amssymb}
\usepackage{amsmath}
\usepackage{natbib}
\usepackage[usenames,dvipsnames]{xcolor}
\usepackage{astrojournals}
\usepackage{multicol}
\usepackage{multirow}
\usepackage{hyperref}
\usepackage{enumerate}
\usepackage{placeins}

\defcitealias{ELVIS}{GK14}

\newcommand{\mvir}{M_\mathrm{v}}
\newcommand{\rvir}{R_\mathrm{v}}
\newcommand{\rmax}{R_\mathrm{max}}

\newcommand{\vmax}{V_\mathrm{max}}
\newcommand{\vpeak}{V_\mathrm{peak}}
\newcommand{\msun}{M_\odot}
\newcommand{\lsun}{L_\odot}
\newcommand{\mstar}{M_{\star}}
\newcommand{\mpeak}{M_\mathrm{peak}}

\newcommand{\mpc}{\mathrm{Mpc}}
\newcommand{\kpc}{\mathrm{kpc}}

\newcommand{\kms}{{\rm km} \, {\rm s}^{-1}}
\newcommand{\lcdm}{$\Lambda$CDM}
\newcommand{\vhalf}{V_{1/2}}
\newcommand{\rhalf}{r_{1/2}}

\newcommand{\vcirc}{V_{\rm circ}}
\newcommand{\sstar}{\sigma_\star}

\title{Too Big to Fail in the Local Group}
\author[S. Garrison-Kimmel et al.]{Shea Garrison-Kimmel$^1$\thanks{$\!$sgarriso@uci.edu},
  Michael Boylan-Kolchin$^{2}$,
  James S. Bullock$^1$, \and
  Evan N. Kirby$^{1,3}$ \\
  \noindent$\!\!$ $^1$Center for Cosmology, Department of Physics and Astronomy,
  University of California, Irvine, CA 92697, USA \\
    \noindent$\!\!$ $^2$Department of Astronomy and Joint Space-Science Institute,
    University of Maryland, College Park, MD 20742-2421, USA \\
  \noindent$\!\!$ $^3$Center for Galaxy Evolution Fellow}
\begin{document}

 \pagerange{\pageref{firstpage}--\pageref{lastpage}} 
 \pubyear{2013}

\maketitle

\label{firstpage}
\begin{abstract} 
We compare the dynamical masses of dwarf galaxies in the Local Group (LG) to 
the predicted masses of halos in the ELVIS suite of \lcdm\ simulations,  a 
sample of 48 Galaxy-size hosts, 24 of which are in paired configuration similar 
to the LG.  We enumerate unaccounted-for dense halos ($\vmax \gtrsim 25~\kms$) in 
these volumes that at some point in their histories were massive enough to 
have formed stars in the presence of an ionizing background ($\vpeak > 30~\kms$). 
Within 300 kpc of the Milky Way, the number of unaccounted-for massive halos 
ranges from 2 -- 25 over our full sample.  Moreover, this ``too big to fail" count 
grows as we extend our comparison to the outer regions of the Local Group: 
within 1.2 Mpc of either giant we find that there are 12-40 unaccounted-for 
massive halos.  This count excludes volumes within 300 kpc of both the MW and 
M31, and thus should be largely unaffected by any baryonically-induced 
environmental processes.  According to abundance matching -- specifically abundance 
matching that reproduces the Local Group stellar mass function -- all of these missing 
massive systems should have been quite bright, with $\mstar > 10^6\msun$.   Finally, 
we use the predicted density structure of outer LG dark matter halos together with 
observed dwarf galaxy masses to derive an $\mstar-\vmax$ relation for LG galaxies that 
are outside the virial regions of either giant.  We find that there is no obvious trend in the relation
over three orders of magnitude in stellar mass (a ``common mass" relation), 
from $\mstar \sim 10^8 - 10^5 ~\msun$, in drastic conflict with the tight relation expected
for halos that are unaffected by reionization.  Solutions to the 
too big to fail problem that rely on ram pressure stripping, tidal effects, 
or statistical flukes appear less likely in the face of these results.
\end{abstract}


\section{Introduction}
\label{sec:intro}
Numerical simulations of structure formation have emerged as a standard 
technique for making and testing predictions of the \lcdm\ model of hierarchical 
galaxy formation \citep{Davis1985,Frenk1988,Warren1992,Gelb1994,Cen1994,
Hernquist1996,Gross1998,Jenkins2001,Wambsganss2004,Springel2005,Boylan-Kolchin2009,Klypin2011}.
These studies have been remarkably successful at reproducing the large-scale properties
of the Universe, but disagreements have periodically emerged on smaller scales.

The smallest dwarf galaxies (stellar mass 
$\mstar \lesssim 10^8\msun$) can be detected and studied best locally,
and thus many of these small-scale problems 
have been identified by comparing observations of Milky Way (MW) satellites 
with subhalos of simulated MW-size hosts.
For example, the ``missing satellites problem" \citep{Kauffmann1993,Klypin1999,Moore1999,Bullock2010}, 
points out that although dark matter (DM)-only simulations predicted a wealth of collapsed 
substructure around the MW, only $\sim10$ bright satellite galaxies are known.  
Though the known count of MW satellites has more than doubled in the past ten years, 
all of these new satellites have been of fairly low mass \citep[e.g.][]{Willman2005,Belokurov2006,Belokurov2007}.
Moreover, even allowing for these new detections in the overall count,
one must still assume that only a small
percentage of subhalos are populated by luminous galaxies in order to explain the 
discrepancy.  It is typical to assume that the brightest ``classical" dwarf spheroidal (dSph) 
galaxies are hosted by the largest subhalos typical of MW-size hosts ($\vmax\sim30~\kms$).

The idea that the most luminous galaxies reside in the most massive halos is
reinforced by the success of the abundance matching (AM) technique, which
accurately reproduces clustering statistics and luminosity functions for $\mstar
> 10^8\msun$ galaxies
\citep{Kravtsov2004,Vale2004,Conroy2006,BehrooziAM,Moster2013}.  Specifically,
AM provides an $\mstar-M_{\rm halo}$ relation by matching DM halo mass functions
from cosmological simulations with stellar mass functions from large-volume
surveys, implicitly assuming that the most luminous galaxies reside in the
largest dark matter halos.  If one extrapolates AM to the dwarf scale, the
resultant satellite stellar mass functions agree well with those of the MW and
M31 satellites for $\mstar\gtrsim10^5\msun$
\citep{Koposov2009,Busha2010,Kravtsov2010,Lunnan2012,MBK2012,Brook2013,ELVIS}.
Below $\mstar \sim 10^5 \msun$, the abundance of galaxies may become more
strongly suppressed than expected in power-law AM extrapolations because the
smallest subhalos ($\vpeak < 30~\kms$) may not have formed stars because of
reionization \citep{Bullock2000,Somerville2002,Sawala2014}.  As discussed in
\citet{ELVIS}, surveys like LSST will test this possibility.

With the advent of the zoom-in technique \citep{Katz1993,Onorbe2013}, 
which focuses the majority of the computational power of a cosmological 
simulation on a small high-resolution region, simulations can now test 
whether these largest subhalos are indeed compatible with the luminous 
MW dSphs, as AM predicts.

\citet{MBK2011,MBK2012} used the zoom-in simulations of the Aquarius 
Suite \citep{Aquarius}, which includes six ultra-high resolution MW-size hosts, 
to compare the internal kinematics of the massive subhalos of MW hosts to the 
brightest MW satellites (those with $\mstar > 10^5\msun$).  They discovered 
that measurements of the stellar velocity dispersions, $\sstar$, indicate 
systematically lower central mass estimates than simulations predict for large 
subhalos~--~that is, the MW dSphs are systematically less dense than the 
subhalos expected to host them, a problem that has been dubbed ``Too Big to Fail" 
(TBTF).  While possibly related to the missing satellites problem, in that the largest
subhalos may not have been found, TBTF is a distinct problem related to the 
internal structure of subhalos, rather than strictly their abundances.  However,
it could be alleviated by the discovery of several new high-density dwarf satellites.   

TBTF may also be tied to the shapes of the inner density profiles of dwarf
halos.  Collisionless simulations predict cuspy central regions, whereas 
measurements by \citet{Walker2011}, \citet{Jardel2012}, \citet{Agnello2012}, 
and \citet{Amorisco2013} indicate cored matter distributions in the larger 
dSphs (Fornax and Sculptor), similar to the cusp-core problem in slightly 
more massive low surface brightness galaxies 
\citep{Flores1994,Moore1994,KuziodeNaray2008,Trachternach2008,deBlok2010,KuziodeNaray2011}.
The slope of the central density profiles are still under debate, 
however~--~\citet{Breddels2013} found that it is unlikely that Fornax, 
Sculptor, Carina, and Sextans are hosted by cored dark matter halos.  The
TBTF problem is independent of the inner slope, however, as it is phrased 
in terms of the integrated mass within the half-light radii of dwarfs, quantities 
that are much more robustly determined observationally than density profile 
slopes.

There have been a number of suggestions proposed for resolving TBTF.  
Some authors have pointed out that self-interactions in the dark matter 
naturally lead to $0.5-1$~kpc cores in dwarf subhalos \citep{Vogelsberger2012,Rocha2013,Elbert2014}, 
though there are  indications that the self-interaction cross section must be velocity 
dependent to satisfy other constraints \citep{Zavala2013}.  Others have investigated 
whether TBTF may be a result of the underlying cosmology of the Aquarius 
simulations, where TBTF was first identified, such as the adopted values of 
$\sigma_8$ and $n_s$ \citep{Polisensky2013} or the assumed coldness of 
the dark matter \citep[][and references therein]{Anderhalden2013,Lovell2013}.
Others have argued that TBTF is a result of the mass of the targeted halos,
pointing to simulations that indicate that smaller hosts, 
$\mvir\sim8\times10^{11}~\msun$, do not typically contain these large, 
dense subhalos \citep{diCintio2011,Wang2012,Vera-Ciro2013}.  It may also 
be that a fraction of the MW-size halos in the Universe do not host these 
dense subhalos \citep{Purcell2012}, though the statistical study of  
\citet{Rodriguez2013} found that the TBTF problem is typical of MW-size
hosts.

Many authors have also noted that TBTF was first identified in collisionless
simulations, which do not account for baryonic forces, and that it is therefore
possible that these missing physics, such as supernova feedback, ram pressure 
stripping, and tidal interactions, may account for the discrepancy 
\citep[e.g.][]{Pontzen2012,Zolotov2012,Arraki2012,BrooksZolotov2012,DelPopolo2012,Brooks2013,Gritschneder2013,Amorisco2013feedback,DelPopolo2014}.  
Although energetic arguments indicate that the former is unlikely in most cases 
\citep{Penarrubia2012,Garrison-Kimmel2013}, there is ample evidence that dwarfs 
are strongly affected by their environment~--~for example, there are only two 
galaxies within 300~kpc of the MW with detected gas (the Magellanic Clouds); 
conversely, there are only two known gas-free field dwarfs within $\sim1~\mpc$ 
of the MW \citep[Cetus and Tucana;][]{Grcevich2009,McConnachie2012}.

Thus far, work on TBTF has focused largely on the subhalos and dSph satellites of the 
MW, while \citet{Tollerud2014} have shown the same issue is seen around 
M31.  To eliminate the uncertain effects introduced by environment, however, one 
should study galaxies beyond the virial radii of the MW and M31, where ram pressure 
and tidal stripping are minimal. Isolated dwarf galaxies in the Local Field (a term we 
will use to refer to the region within $1.2~\mpc$ of either the MW or M31, but more 
than $300~\kpc$ from both) do not appear to be denser than the MW dSphs 
\citep{Kirby2013}, but predictions for halo properties in the Local Field have thus far 
been sparse.  

In this paper, we examine both satellite and field dwarf halos around the 
hosts of the Exploring the Local Volume in Simulations (ELVIS) Suite
\citep[][hereafter GK14]{ELVIS}, a set of zoom-in simulations focused on 
LG-like environments that resolve $\sim3~\mpc$ regions without 
contamination from low resolution particles, for the TBTF problem.  Specifically, 
we count the number of  ``massive failures"~--~large halos 
($\vpeak > 30~\kms$) that do not have luminous counterparts~--~both within 
$300~\kpc$ of the 48 MW-size hosts and in the fields surrounding the 
LG analogs.  Because the ELVIS Suite adopts cosmological parameters from the 
WMAP-7 results \citep[$\sigma_8 = 0.801$, $\Omega_m = 0.266$, 
$\Omega_\Lambda = 0.734$, $n_s = 0.963$, and $h = 0.71$;][]{Larson2011}, 
which includes a significantly lower value of $\sigma_8$ than the WMAP-1 parameter set 
adopted for the Aquarius simulations, we will also test whether an updated cosmology 
alleviates the problem.  As we show below, however, we predict that there are many such
unaccounted-for dense halos throughout the Local Volume.  If these halos preferentially
host low-luminosity or low-surface brightness galaxies, then future surveys may detect
them.  

This paper is organized as follows. In \S\ref{sec:sims}, we briefly describe the
simulations and analysis pipeline used in this work.  In \S\ref{sec:rmaxvmax},
we present empirical scaling relations between the structural parameters of
subhalos and field halos and explicitly compare the properties of small halos
near isolated hosts with those in paired environments.  \S\ref{sec:fails}
presents the counts of massive failures around each host both within $300~\kpc$
of each host (\S\ref{ssec:rvirfails}) and in the field surrounding the Local
Group analogs (\S\ref{sssec:fieldfails}), as well as a discussion of
incompleteness (\S\ref{sssec:incompleteness}).  We conclude with an analysis of
the relationship between $\mstar$ and $\vmax$ for the known dwarfs in the Local
Field in \S\ref{sssec:mstarvmax}.  Our results are summarized in \S\ref{sec:conclusions}.

\section{Simulations:  The ELVIS Suite}
\label{sec:sims}	
The simulations used in this work, the ELVIS Suite, are described in detail 
in \citetalias{ELVIS}.  The large scale properties of the LG analogs and the 
individual properties of the paired and isolated halos (along with their 
identifying names) are given in that work.  Here we briefly summarize the 
simulations and the analysis pipeline used in this paper.

The suite is comprised of $36$ collisionless simulations, half of which are 
focused on a pair of dark matter halos whose masses, relative kinematics, 
and environments are similar to the dark matter halos that host the MW
and Andromeda (M31) galaxies.  The remaining twenty-four simulations are 
focused on isolated halos that are mass-matched to those in the pairs.  Because the
mass estimates for the MW and M31 agree within errors \citep{Marel2012,Boylan-Kolchin2013}, 
both hosts in each paired simulation may separately be considered as an MW 
analog; the ELVIS Suite therefore contains a total of $48$ MW-size systems.  The
distribution of virial masses\footnote{Throughout, we define $\mvir$ as the mass 
within a sphere of radius $\rvir$ that corresponds to an over density of 97 relative 
to the critical density.} $\mvir$ of the ELVIS hosts nearly evenly samples 
the mass range between $10^{12}\msun$ and $2.85\times10^{12}\msun$.  
All halos in the suite were simulated with a $z=0$ Plummer equivalent force softening of 
$141$~pc in the high resolution region, which contains particles with a mass 
$m_\mathrm{p} = 1.89\times10^5\msun$.  Additionally, three of the isolated 
hosts were re-simulated with a factor of $2^3$ more particles 
($m_\mathrm{p} = 2.4\times10^4\msun$) in the high-resolution region 
and a corresponding $z = 0$ softening length of $70$~pc.  We use these runs to 
demonstrate the convergence of subhalo structural parameters in Appendix~A.

Bound substructures are identified with \texttt{Rockstar}, a six dimensional
friend-of-friends halo finder \citep{rockstar}.  For this analysis, the
relevant properties are $\vmax$,  the maximum of the circular velocity profile, 
and $\rmax$, the radius at which the circular velocity peaks.  We additionally 
select halos that are expected to have formed stars based upon $\vpeak$, which 
is defined as $\vmax$ of the main branch of the halo's merger tree, built 
with \texttt{Consistent Trees} \citep{Behroozi2013b}, at the timestep when 
the halo reaches its maximal mass \citetext{see \citetalias{ELVIS} for more details}.  

Each run in the ELVIS Suite was initialized with a large high-resolution region 
to specifically enable study beyond the virial radius of the giant halos without 
contamination due to low resolution (high mass) particles.  Specifically, only 
four (Thelma \& Louise, Sonny \& Cher, Hall \& Oates, and Siegfried \& Roy) of 
the twelve LG realizations contain such contaminating particles within 
$1.2~\mpc$ of either halo center.  In those cases, moreover, the contamination
is minimal:  within $1.2~\mpc$ of either halo center, the contamination by mass
is only $0.06\%$, $0.01\%$, $0.007\%$, and $0.0008\%$, respectively.  
In addition, the nearest low resolution particles in these four systems are quite
distant:  $0.8~\mpc$, $0.97~\mpc$, $1.01~\mpc$, and $1.09~\mpc$.  Catalogs of 
halos in the fields around the ELVIS hosts are therefore complete and nearly 
entirely free of contamination at much larger distances than previous 
high-resolution simulations (the well known CLUES project, \citealt{Gottloeber2010},
and recent work by \citealt{Sawala2014}, are notable exceptions).

The goal of this work is to compare predicted halo densities to those of LG dwarfs 
at scales comparable to their observed half-light radii ($\sim200 - 1000$~pc).  
Because our fiducial set of simulations lacks the resolution required make direct 
predictions at scales below $\sim1000$~pc, we instead use the well-converged
structural parameters ($\vmax$ and $\rmax$) together with several reasonable 
choices for analytic profiles in order to extrapolate to the scales of observed dwarfs. 

$\rmax$ and $\vmax$ together uniquely define a Navarro-Frenk-White 
\citep[NFW;][]{NFW} profile:
\begin{equation}
\rho(r) = \rho_0\left(\frac{2.1626\,r}{\rmax}\right)^{-1}\left(1+\frac{2.1626\,r}{\rmax}\right)^{-2},
\label{eqn:NFW}
\end{equation}
where $\rho_0$ is defined such that the mass within $\rvir$ is equal to $\mvir$.
For a given shape parameter $\alpha$, one may also calculate a unique Einasto
profile \citep{Einasto} based upon $\rmax$ and $\vmax$, though the scalings
between the characteristic radius $r_{-2}$ and $\rmax$ and between 
$\rho_{-2}$, the density at $r_{-2}$, and $\vmax$ depend upon the 
shape parameter:
\begin{equation}
\rho(r) = \rho_{-2}\exp\left(-\frac{2}{\alpha}\left[\left(\frac{A(\alpha)\,r}{\rmax}\right)^{\alpha}-1\right]\right),
\label{eqn:Einastodensity}
\end{equation}
where $r_{-2} = \rmax/A(\alpha)$.  Appendix~B defines $A(\alpha)$ and explicitly 
compares the NFW and Einasto profiles.

As mentioned above, in addition to the forty-eight halos simulated at the fiducial 
resolution, the ELVIS Suite also contains high-resolution re-simulations of three 
of the isolated hosts.  We use these halos to ensure the convergence of $\vmax$ 
and $\rmax$ (see Appendix~A) and find that a power law fit to the $\rmax-\vmax$ 
relationship,
\begin{equation}
\left(\frac{\rmax}{\rm 1~kpc}\right) = A\left(\frac{\vmax}{\rm 10~\kms}\right)^{1.47},
\label{eqn:310fit}
\end{equation}
describes both populations well.  For $\vmax > 15~\kms$ and $\rmax > 0.5~\kpc$,
the normalizations, $A$, differ by less than 3\%.

Therefore, although the standard ELVIS runs lack the resolving power to determine
inner differential density profiles, the integral properties of the halos of interest are
well constrained.  As pointed out by \citet{diCintio2013}, however, the number of 
massive failures is dependent on the individual subhalo density profiles.  We 
therefore investigate both the NFW profile and a range of Einasto profiles.  We
primarily present results with $\alpha = 0.18$, which \citet{Aquarius} showed 
is generally a slightly better fit to subhalos in ultra-high resolution DM-only 
simulations than an NFW profile, but also use $\alpha = 0.15$ as an example of a 
peaky Einasto profile and $\alpha=0.28$ to sample flatter density profiles 
(this range also encompasses the results of \citet{Gao2008} and \citet{Navarro2010},
though both works investigated more massive halos).  We will see below that, 
while exact numbers may depend strongly on the assumed density profile, our overall 
conclusions hold for all profiles in this regime.

\section{$\mathbf{\rmax-\vmax}$ Relationships}
\label{sec:rmaxvmax}
As stated above, the parameters $\rmax$ and $\vmax$, plus an
assumed functional form for the density profile, fully define 
the circular velocity curve of a halo.  The relationship between
these parameters is therefore fundamental to the TBTF problem.  
In this section, we present fits to $\rmax$ as a function of $\vmax$ 
and compare the paired and isolated samples to search for biases in 
the structure of dwarf halos related to the environments of their hosts.

\subsection{Subhalo scaling relations within 300~kpc}
\label{ssec:virstucture}
\begin{figure}
\centering
\includegraphics[width=0.49\textwidth]{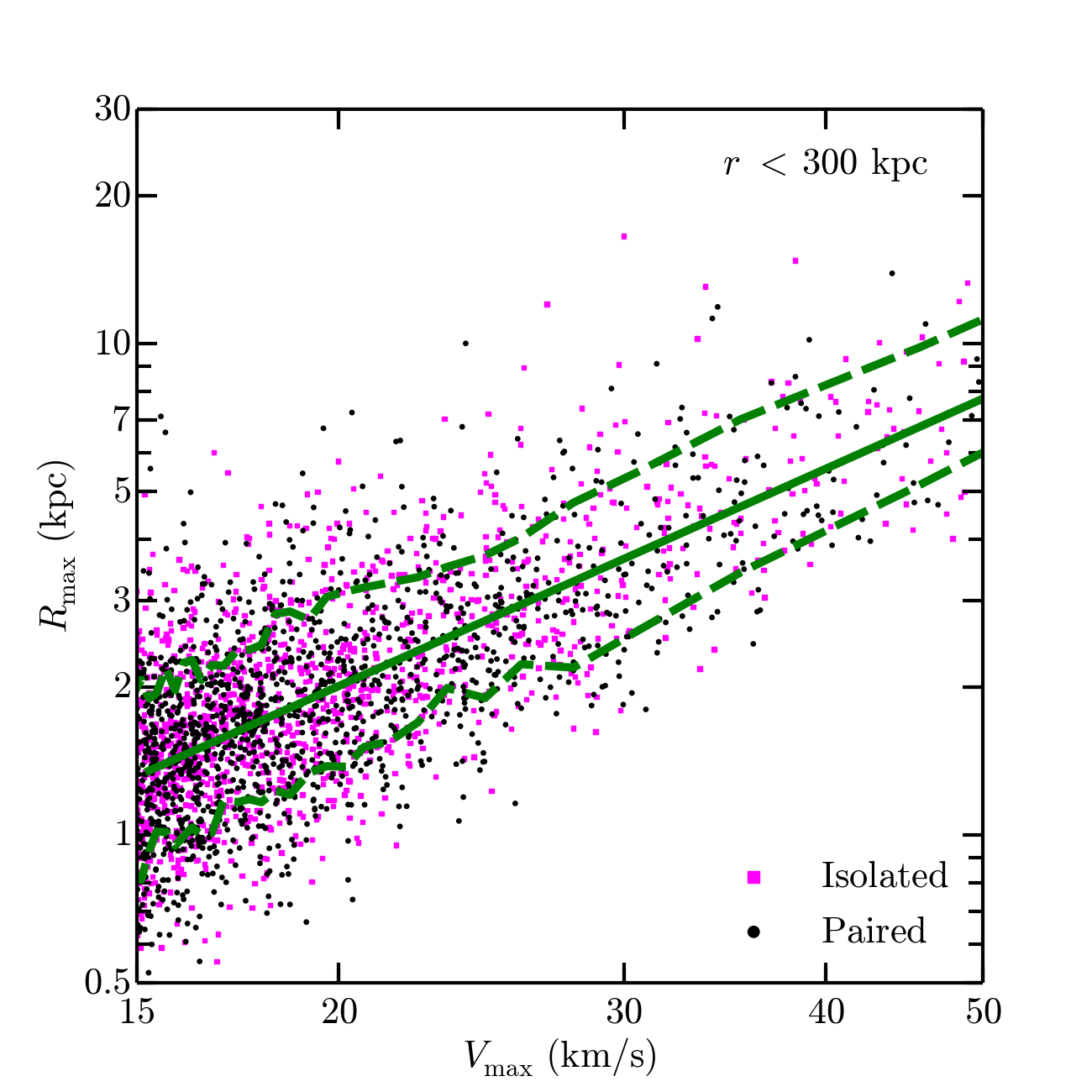}
\caption{The relationship between $\rmax$ and $\vmax$ for subhalos in
the ELVIS Suite within 300~kpc of each host.  Subhalos near the paired 
hosts are plotted as black circles; those near isolated hosts are indicated 
by magenta squares.  The thick green line plots the fit to all the halos and 
the dotted green lines encompass $68\%$ of the points; the fits to these 
relations and the isolated and paired populations separately are given in
Table~\ref{tab:300norms}.  As the two datasets follow nearly identical 
relations and have consistent mass functions within the virial radii \citepalias{ELVIS}, 
we will combine the samples for better statistics when counting discrepant halos
within $300~\kpc$ of the hosts.}
\label{fig:rv300kpc}
\end{figure}

\begin{table}
\centering
\begin{tabular}{cccc}
Sample & $A_{\rm fit}$ & $A_{+68\%}$ & $A_{-68\%}$ \\ \hline \hline
Isolated &  0.747 & 1.09 & 0.521 \\ 
Paired & 0.704 & 1.00 & 0.499 \\
Combined & 0.725 & 1.06 & 0.511 \\ \hline
\end{tabular} 
\caption{Fit results for the $\rmax-\vmax$ relationship defined in 
Equation~\ref{eqn:310fit}.  Listed are the normalizations resulting 
from fitting the data (Column~1) and from fitting the 68\% scatter 
about that relation in bins of 100 points (Columns~2~and~3), 
separately for subhalos ($r<300~\kpc$) of the isolated and paired 
hosts, and when combining the datasets (the green lines in 
Figure~\ref{fig:rv300kpc}).}
\label{tab:300norms}
\end{table}

Though the ELVIS Suite contains $48$ MW-size halos, only those in the paired 
sample are truly fair comparisons to the MW.  However, \citetalias{ELVIS} showed 
that subhalo counts at fixed mass are identical between the two samples (when 
controlling for the host mass); we therefore begin by comparing the structural 
properties of subhalos of isolated and paired hosts to determine if the samples 
may be combined when counting massive failures within $300~\kpc$ of the hosts.

Figure~\ref{fig:rv300kpc} plots the relationship between $\rmax$ and $\vmax$ 
for all subhalos within 300~kpc of the ELVIS hosts.  Subhalos of the isolated 
hosts are plotted as magenta squares and those of hosts in LGs are indicated 
by black circles.  The green line plots a fit to all the subhalos, holding the slope
fixed to that in Equation~\ref{eqn:310fit}; the dashed lines indicate the 68\% 
scatter about that relation, calculated in running bins of 100 subhalos.  The normalization 
of the fit, along with that of fits to the scatter above and below the relation, are 
given in Table~\ref{tab:300norms} separately for the two populations, which 
differ only at the 5\% level, and when combining the datasets.  Any variance 
between subhalos of isolated and paired halos is well within the intrinsic scatter,
and we therefore perform the remainder of our analysis within $300~\kpc$ using 
subhalos of both isolated and paired hosts to maximize our statistics.  

Because the subhalo properties in the paired and isolated system agree, we find no 
evidence that the results of \citet{MBK2011,MBK2012} are affected by their study 
of isolated hosts.  However, at the typical size of a TBTF halo 
($\vmax\sim30-50~\kms$), the median $\rmax$ of a subhalo in the ELVIS systems
is $25\%-30\%$ larger than those in the Aquarius simulations, consistent with
the offset in $\sigma_8$ \citep{Zentner2003,Polisensky2013}.  This allows each 
dwarf to live in more massive hosts, and will lead to fewer discrepant halos.  
We will discuss this further in Section~\ref{ssec:rvirfails}.

\subsection{Halo scaling relations in the Local Field}
\label{ssec:fieldstructure}
\citetalias{ELVIS} showed that there are systematic differences between 
the environments surrounding isolated and paired halos, but did not compare 
the internal structure of halos in each environment.  We therefore search 
for biases in the Local Field (LF) related to the larger-scale environments by 
comparing the relationship between $\rmax$ and $\vmax$ for field halos 
around isolated MWs and those in LGs.

\begin{figure}
\centering
\includegraphics[width=0.49\textwidth]{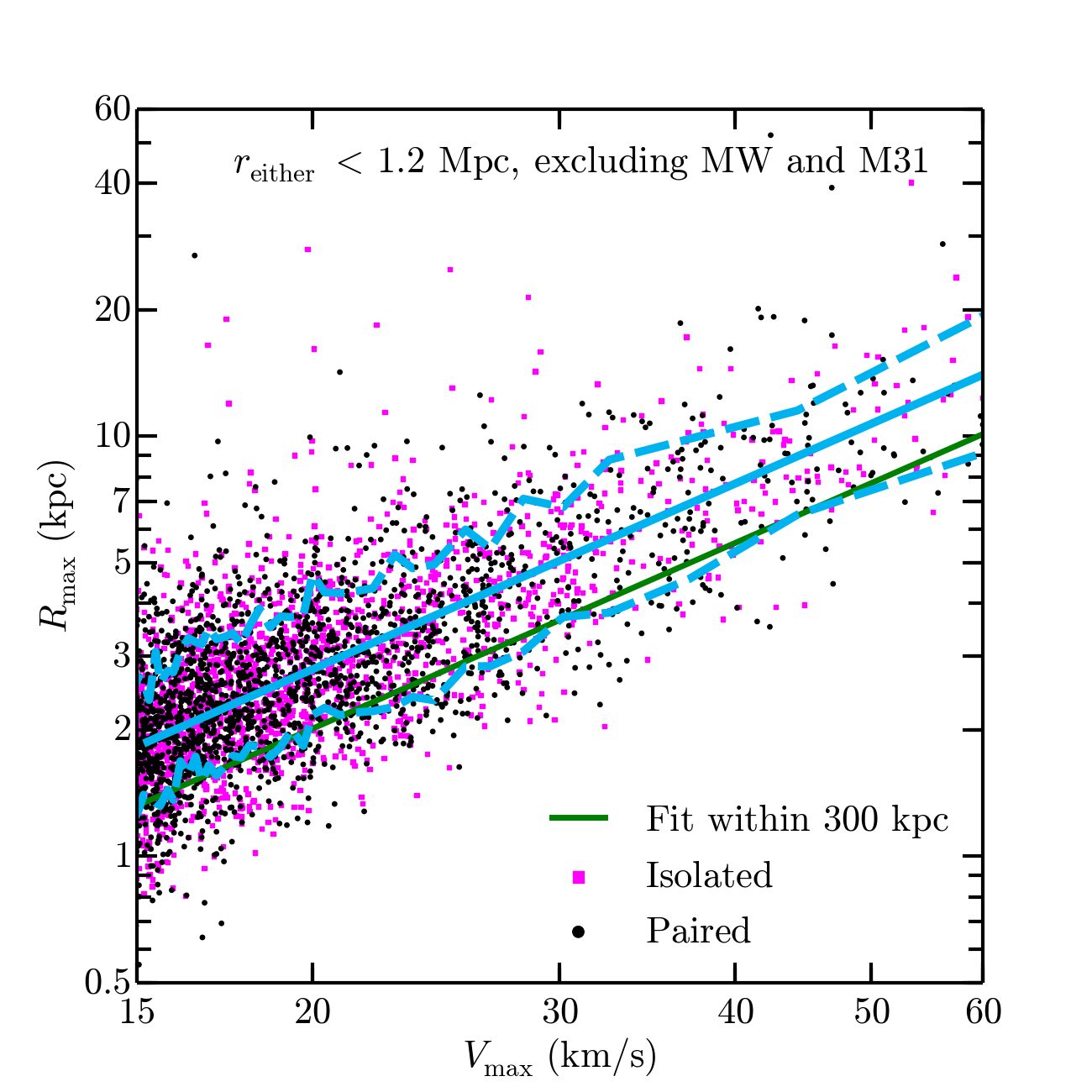}
\caption{Identical to Figure~\ref{fig:rv300kpc}, but plotting the relationship
between $\rmax$ and $\vmax$ of halos that reside in the Local Field~--~the 
region within 1.2~Mpc of either host, but more than 300~kpc from both giants.  
The cyan line plots a power-law fit to all the halos with a log slope held equal to 
that in Equation~\ref{eqn:310fit}; the normalization for all the data and for the 
individual datasets, along with fits to the scatter (dashed lines) are given in 
Table~\ref{tab:fieldnorms}.  The green line plots the fit within $300~\kpc$, 
where halos are systematically denser at fixed $\vmax$ due to tidal stripping.  
}
\label{fig:rvfield}
\end{figure}

Figure~\ref{fig:rvfield} plots this relationship in the LF (the region within
$1.2~\mpc$ of either giant, but more than 300~kpc from both).  The relation is
again well fit by a power law with a log slope of $1.47$
(Equation~\ref{eqn:310fit}); such a fit is plotted as a light blue line and the
68\% scatter about that fit, again calculated in running bins of 100 halos, is
indicated by the dashed lines.  As expected from tidal stripping arguments
\citep[see][]{Zentner2003,Kazantzidis04,Diemand2007}, the average densities of
field halos are significantly lower than subhalos at fixed $\vmax$, as can be
seen from the green line, which indicates the fit within $300~\kpc$ plotted in
Figure~\ref{fig:rv300kpc}.  We again fix the slope of the fits and find the
normalizations given in Table~\ref{tab:fieldnorms}.

Although the normalizations presented in Table~\ref{tab:fieldnorms}
for the isolated and paired samples agree at the percent level,
\citetalias{ELVIS} showed that the number counts do not agree beyond the 
virial radius of each host.  As we are explicitly concerned with both the 
number and structure of field halos, we will use only those surrounding 
the paired hosts to count massive failures in the LF.  Moreover, as in 
\citetalias{ELVIS}, we will exclude the two systems with a third large halo 
in the Local Volume (Siegfried \& Roy and Serena \& Venus) when studying
the LF. However, the apparent lack of structural differences indicates 
that detailed ultra-high resolution simulations of isolated dwarf galaxies 
in the field should be accurate analogs to Local Field dwarfs that have not 
yet interacted with either giant.

\begin{table}
\centering
\begin{tabular}{cccc}
Sample & $A_{\rm fit}$ & $A_{+68\%}$ & $A_{-68\%}$ \\ \hline \hline
Isolated &  1.016 & 1.443 & 0.723 \\ 
Paired & 0.994 & 1.437 & 0.709  \\
Combined & 1.005 & 1.448 & 0.719 \\ \hline
\end{tabular}
\caption{The normalizations for the $\rmax$-$\vmax$ relationship 
(Equation~\ref{eqn:310fit}) in the Local Field as well as fits to 
the envelope that contains 68\% of the halos, as in 
Table~\ref{tab:300norms}.  For the paired systems, the Local Field 
is defined as the region within $1.2~\mpc$ of either host, but 
excluding all subhalos within $300~\kpc$ of both hosts; the isolated
``Local Fields" include all halos within $1.2~\mpc$ of the main host
only, again excluding all subhalos within $300~\kpc$.}
\label{tab:fieldnorms}
\end{table}

\section{Massive Failures in the ELVIS Suite}
\label{sec:fails}

\subsection{Counting massive failures within 300~kpc}
\label{ssec:rvirfails}
Qualitatively, we are concerned with counting halos that are massive 
enough that they should have formed stars, but that have no obvious 
luminous counterparts in the local Universe.  We select halos with 
$\vpeak > 30~\kms$ as ``massive enough" because halos larger
than 30 km/s should be able to retain substantial gas in the presence 
of an ionizing background and therefore, \textit{in principle}, should 
form stars \citep{Babul1992,Efstathiou1992,Thoul1996,Gnedin2000,Okamoto2008};
however, we must also carefully define the criteria to be a ``luminous
counterpart" of a galaxy in our sample.   In what follows, we describe 
two ways of counting subhalos that have no obvious luminous counterparts.

As in \citet{MBK2011}, our observational sample is comprised of the 
satellites within $300~\kpc$ of the MW with $\mstar>2\times10^5\msun$, 
excluding the Sagittarius dwarf and the Magellanic Clouds.  
Sagittarius is currently undergoing an interaction with the MW disk 
and is therefore likely not in equilibrium; the dwarf irregular 
Magellanic Clouds are removed from the sample because satellites as 
large as the Magellanic Clouds are rare around MW-size hosts 
\citep{Boylan-Kolchin2010,Busha2011,Tollerud2011}, and therefore do 
not have corresponding subhalos in many of the ELVIS systems.
Our observational sample is thus likewise comprised of nine galaxies 
with $L>10^5\lsun$:  the classical dSphs and Canes Venatici (CVnI).  

We now turn to the problem of assigning galaxies to subhalos,
and identifying subhalos without luminous counterparts.  The 
original formulation of TBTF counted unidentified subhalos as 
objects with circular velocity profiles that were at least $2\sigma$
above the observed circular velocity of each dwarf at its half-light 
radius ($\vhalf = \vcirc(r = \rhalf)$).  These subhalos clearly lack 
observational counterparts.  We will adopt a similar counting procedure, 
but instead use $1\sigma$ errors to define over-dense outliers.  
Specifically, we will refer to subhalos with $\vpeak > 30~\kms$
that are more than $1\sigma$ denser (at $\rhalf$) than any of the 
MW dwarfs as "strong massive failures".

This ``strong massive failure" formulation, which mirrors that 
originally used in \citet{MBK2011,MBK2012}, is particularly conservative.  
By counting only subhalos that are denser than all of the MW dwarfs, it 
ignores the potentially large number of subhalos that are consistent with 
hosting only the densest observed dwarfs.   Most MW-size hosts contain 
several subhalos that can \textit{only} host either Draco or Ursa Minor, 
but nothing else.  Since clearly only one halo can actually host Draco, 
this way of counting under-estimates the magnitude of the problem.  Moreover, 
the ``strong massive failure" definition is highly dependent on a single 
object, the densest MW dSph (Draco).  If Draco did not exist, the strong 
massive failure count would be much larger.  Similarly, if Draco were twice 
as dense, the strong massive failure count would approach zero.  Ideally, 
we would like to find a measure that is less sensitive to the properties 
of a single object.

With these issues in mind, we introduce a second way of counting unidentified
massive subhalos, which we refer to as the ``massive failure" count.  These are
halos that were massive at infall (with $\vpeak > 30~\kms$) and that have no 
observational counterpart after each dense satellite is assigned to a single subhalo.
Specifically, we find all halos that are \textit{at least} as dense as Draco and Ursa 
Minor (in practice this demands that today halos have $\vmax \gtrsim 25~\kms$).  
We then examine the subset that are consistent with either Ursa Minor or Draco 
and remove the most massive possible counterpart to those galaxies.  The remaining 
set allows us to enumerate unaccounted-for, yet massive, halos.  We will discuss the
impact of selecting Draco and Ursa Minor for this process below.

To summarize, we will count two classes of discrepant halos in the ELVIS
Suite. \textbf{Strong massive failures} are too dense to host any of the
bright MW dSphs, with circular velocities at $\rhalf$ that are above
the $1\sigma$ constraints for all the dwarfs in the sample.  \textbf{Massive
failures} include all strong massive failures plus all 
massive halos that have densities consistent with the high density dwarfs (Draco and Ursa Minor) but
that can't be associated with them without allowing
a single galaxy to be hosted by multiple halos.   For typical
profiles, subhalos with $\vmax \lesssim 25-30~\kms$ can host a low density 
dwarf, and thus are never selected as a massive failure; the {\bf massive failures}
are therefore generally subhalos that started out dense ($\vpeak > 30~\kms$) and
remain dense ($\vmax \gtrsim 25~\kms$) at $z = 0$.

\begin{figure}
\centering
\includegraphics[width=0.49\textwidth]{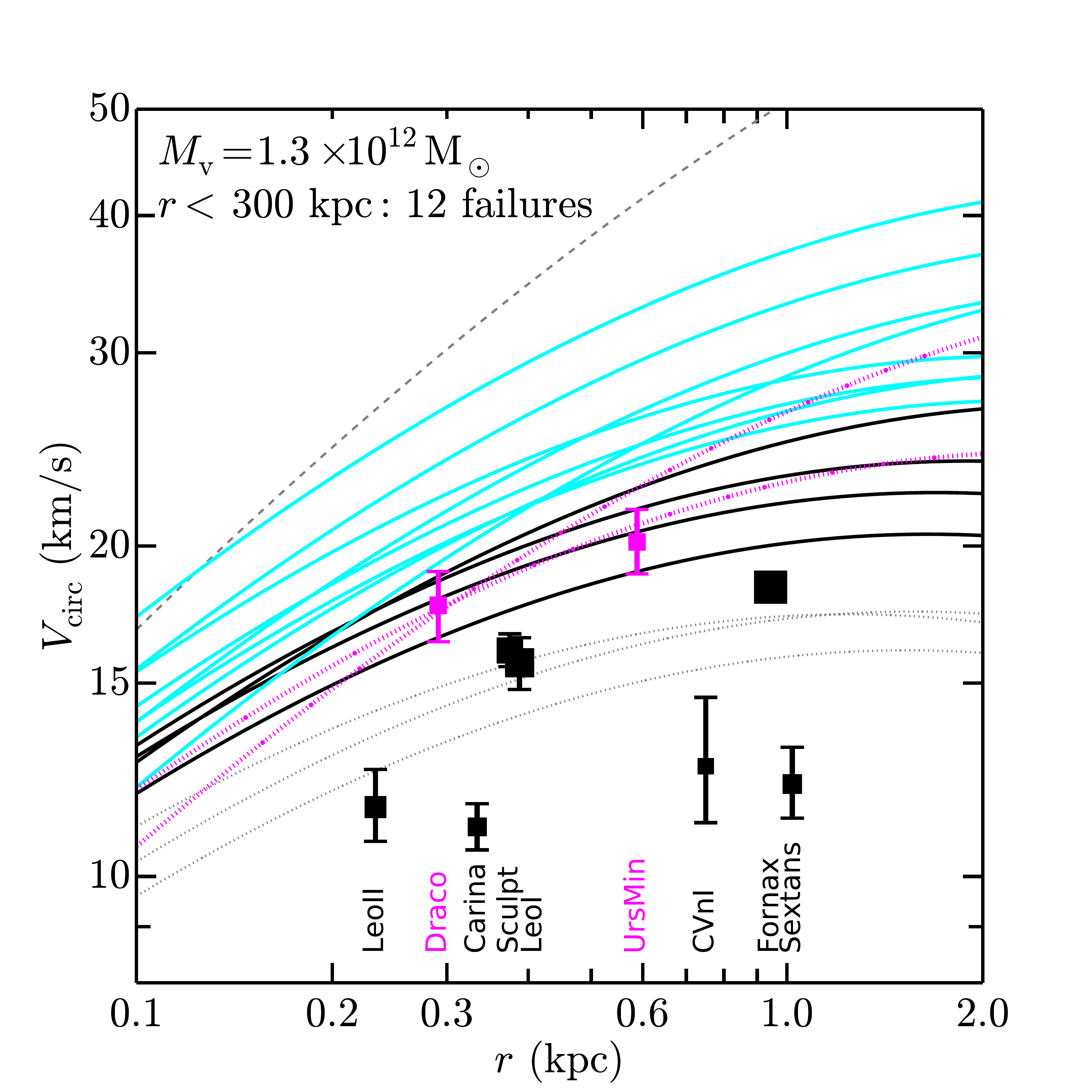}
\caption{Rotation curves, assuming Einasto profiles with 
$\alpha=0.18$, of all resolved halos with $\vpeak > 30~\kms$ within 
300~kpc of the center of Douglas (based on measured $\vmax$ and $\rmax$ values in the simulation).  
Plotted as black points are the data for 
the MW satellites brighter than $2\times10^5$~L$_\odot$ compiled in 
\citet{Wolf2010}, with sizes proportional to the log of their stellar masses.  
The cyan lines indicate strong massive failures~--~subhalos that are too 
dense to host \emph{any} of the MW dSphs.  The black lines plot the 
additional subhalos that are identified as massive failures according to 
the stricter definition given in the text:  halos with $\vpeak > 30~\kms$ 
that are not accounted for by the dense galaxies in the observational sample.  
The subhalos with $\vpeak > 30~\kms$ that are selected to host the high
density galaxies, Draco and Ursa Minor, are indicated by dotted magenta 
lines, with their associated galaxies plotted as magenta squares.  The 
dotted lines plot the subhalos that are consistent with at least one of the
remaining seven dwarfs in our sample, which are allowed to reside in
multiple such subhalos.  The grey dashed line indicates the sole subhalo 
of Douglas expected to host a Magellanic Cloud ($\vmax > 60~\kms$), 
which we exclude from our analysis.  Not plotted are 40 resolved
($\vmax > 15~\kms$) subhalos with $\vpeak<30~\kms$.  In all, 
Douglas hosts twelve unaccounted-for massive failures, including eight 
strong massive failures that are too dense to host any bright MW dSph.}
\label{fig:Rvircurves}
\end{figure}

Figure~\ref{fig:Rvircurves} provides an illustration of these definitions.
Shown are rotation curves of all $\vpeak > 30~\kms$ halos identified within
$300~\kpc$ of an $\mvir = 1.3\times10^{12}~\msun$ halo (Douglas, a paired host
in the ELVIS sample).  The solid black lines and solid cyan lines plot massive
failures; the latter are {\em strong} massive failures because they are denser
than every dwarf.  The dotted curves indicate subhalos that had $\vpeak >
30~\kms$ but that are {\em not} massive failures~--~the magenta dotted lines are
those selected to host Draco and Ursa Minor, and the grey dotted lines plot
systems that have been stripped enough to host the lower density galaxies at $z
= 0$.  The curves correspond to Einasto profiles with $\alpha = 0.18$, normalized 
using the measured $\rmax$ and $\vmax$ values for each identified system.  The 
dashed grey line indicates the lone Magellanic Cloud analog in Douglas, defined 
as subhalos with present day $\vmax > 60~\kms$ \citep{Stanimirovic2004}, 
which is eliminated from our analysis.  Our cut is again less conservative 
than that in \citet{MBK2011}: the criterion used by those authors would 
eliminate approximately one additional subhalo per host, on average (i.e. 
they would measure one fewer strong massive failure per host).

\begin{figure*}
\centering
\includegraphics[width=0.98\textwidth]{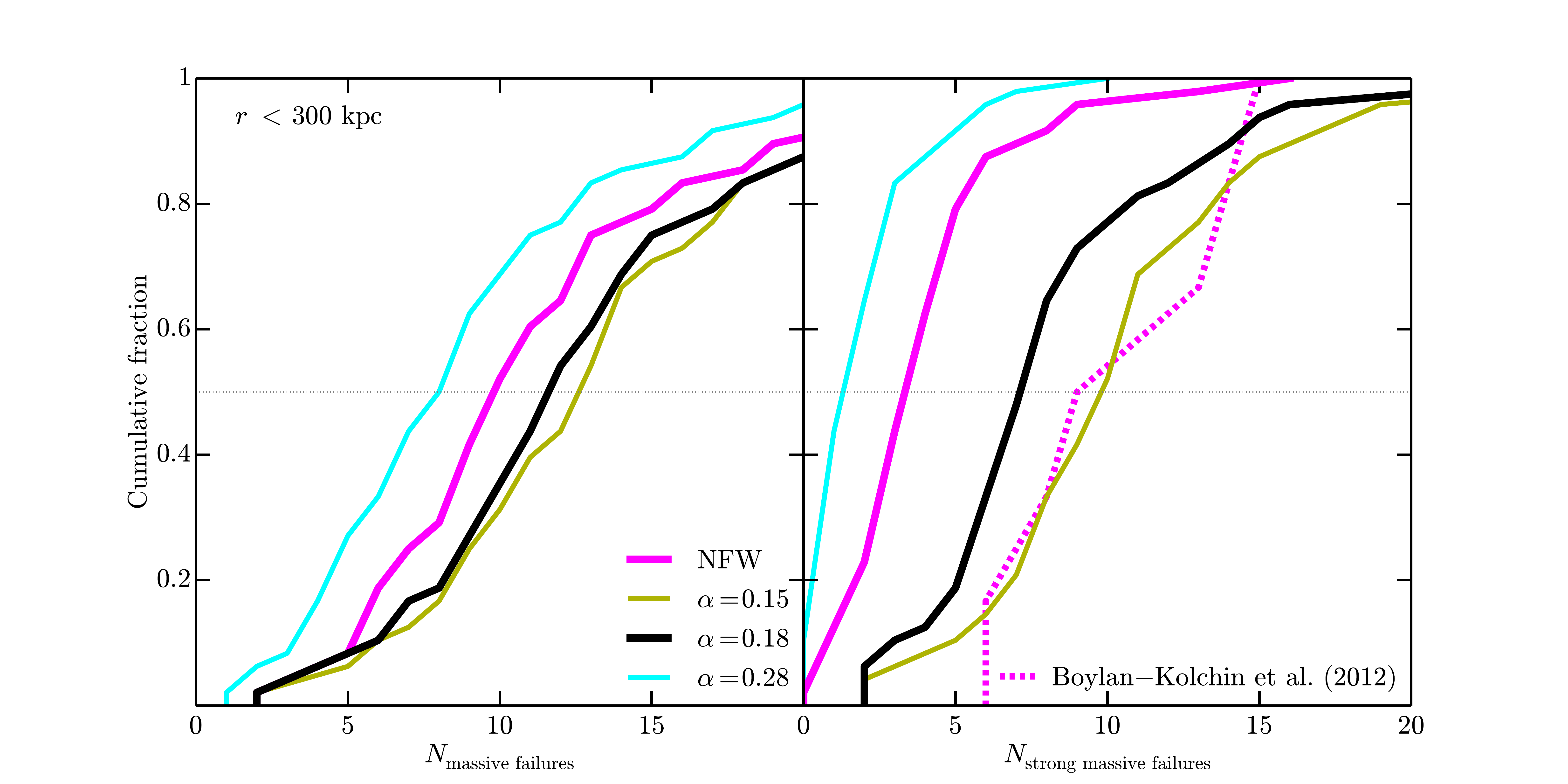}
\caption{The fraction of hosts (out of 48) with fewer than $N$ massive failures 
on the left and $N$ strong massive failures on the right within $300~\kpc$ of each 
host, as a function of $N$.  Plotted are results assuming an NFW density profile 
(magenta) and Einasto profiles with $\alpha = 0.15$ (dark yellow), $0.18$ (black), 
and $0.28$ (cyan).  In the left panel, we plot the number of strong massive failures in 
the Aquarius hosts as a dashed magenta line.  Less than $10\%$ of the ELVIS hosts 
contain no strong massive failures and we predict $\sim12$ massive failures within 
$300~\kpc$ of the MW.}
\label{fig:cumhist310}
\end{figure*}

The data points in Figure~\ref{fig:Rvircurves} indicate measurements of 
$\vhalf$ at $\rhalf$ for the MW dSphs in our sample (taken from 
\citealt{Wolf2010}, who used data from \citealt{Walker2009} along with 
data from \citealt{Munoz2005,Koch2007,Simon2007} and \citealt{Mateo2008}).
\footnote{For simplicity, we exclude galaxies within $300~\kpc$ of M31~--~many 
of the M31 satellites have substantial contributions from baryons within
$\rhalf$, making a measurement of the central dark matter density very 
difficult.  However, the central masses of the M31 dSphs appear to be 
consistent with the MW dSphs \citep{Tollerud2012}, and are therefore
inconsistent with the subhalos expected to host them \citep{Tollerud2014}.}
The points are sized by the log of the stellar mass of each galaxy.  Plotted 
in black are the low density MW dSph galaxies.  The magenta points indicate 
the high density dSphs, Draco and Ursa Minor, which may only be associated 
with a single subhalo in each host (indicated by the dotted magenta lines) 
when counting massive failures.  If the data points for Draco or 
Ursa Minor were $10~\kms$ higher, the strong massive failures (cyan lines) 
would vanish but the number of massive failures (cyan and black lines) would remain 
unchanged.

Figure~\ref{fig:cumhist310} summarizes the results of counting massive failures in
the complete set of forty-eight hosts, where each line corresponds to a different 
assumed density profile shape.  Black lines show results for our fiducial choice, 
an $\alpha = 0.18$ Einasto profile; also shown are the implied distributions for 
NFW profiles (magenta), an underdense Einasto (cyan; $\alpha = 0.28$), and an 
overdense Einasto (dark yellow, $\alpha = 0.15$).  The left panel indicates the 
cumulative distribution of massive failures and the right plots the same for 
strong massive failures; also plotted as a dashed magenta line is the distribution 
of $1\sigma$ discrepant subhalos from the Aquarius simulations,
which we discuss below.  As explained above, the strong 
definition is highly sensitive to the densest dwarf; it is likewise strongly
dependent on the density profile, with medians varying between $2$ and $10$ for 
those chosen here.  The number of massive failures, however, is more consistent 
and varies by a maximum of $\sim5$~--~the median varies from $8.5$ for 
$\alpha=0.28$ to $13$ for $\alpha=0.15$.

All of the forty-eight hosts contain at least two strong 
massive failures for $\alpha = 0.18$; using the slightly less dense NFW profile 
results in only one (iHera, with $\mvir = 1.22\times10^{12}\msun$) of the 
forty-eight hosts (2\%) containing no strong massive failures.  Even the least 
dense profile considered here ($\alpha = 0.28$) leads to only five hosts (10\%) 
with no strong failures. \footnote{For completeness sake, we note that the massive 
failures are drastically reduced in number or disappear completely if we assume a 
strongly cored or flat inner profile ($\alpha = 0.5-1$).}   These results are similar 
to the expectations of \citet{Purcell2012}, who estimated the prevalence of strong 
massive failures in Milky-Way size hosts using a semi-analytic formalism,  though in 
detail we have found slightly higher fractions of systems with strong massive failures.

The problem is revealed as more serious when we enumerate all unaccounted-for
massive halos, however.  None of the ELVIS hosts are without massive failures:
the least problematic MW analogs host $\sim3$ dense subhalos without bright
counterparts~--~more than twice the number of known dense satellites.  Unless
the spatial distribution of dense satellites is highly anisotropic such that
their on-the-sky density drastically increases behind the plane of the disk, it
is unlikely that this disagreement can be reconciled via incompleteness
arguments.  However, one explanation of the observed lack of bright satellites
between $100-400~\kpc$ of the MW \citep{Yniguez2013} is that there are as many
as $\sim10$ missing MW satellites with $L>10^5\lsun$~--~TBTF
may be explained if the majority of these missing galaxies are as dense or 
denser than Draco, though there is no a priori reason to believe this to be 
the case.

The choice of Draco and Ursa Minor as our high-density dwarfs is based on the
observation that they are the only two systems that demand to be hosted by
$\vmax > 20~\kms$ halos to high significance.  Nevertheless, it is useful
to investigate how our massive failure count would change if we altered
this choice.   The number of massive failures shrinks if only Draco
or only Ursa Minor is selected to be uniquely hosted (the medians vary 
between $5-11$ for Draco only and $6-11$ for Ursa Minor only), but adding 
more dSphs to this list identifies only a few more subhalos as massive failures:  
including the three additional galaxies with $\vhalf > 15~\kms$  (Fornax, Leo~I, 
and Sculptor) raises the median per host to only $11-13$.  That is, there are 
$\sim10$ subhalos per host as dense or denser than Draco and Ursa Minor, but 
there are only $\lesssim4$ additional subhalos with central densities similar to 
Fornax, Leo~I, and Sculptor that have reached $\vpeak \geq 30~\kms$.

Our results are consistent with the expectation that lowering $\sigma_8$ helps
to alleviate TBTF.  The distribution of the number of strong massive failures in
the Aquarius hosts is plotted as the dotted magenta line in
Figure~\ref{fig:cumhist310}.  As in \citet{MBK2012}, NFW profiles have been
assumed in the inner region of the halos.  Though the sample size is much
smaller ($6$ instead of $48$), there are significantly more massive failures in
the WMAP-1 cosmology than result from the updated WMAP-7 values, in agreement
with \citet{Lovell2013} and \citet{Polisensky2013}.  Note, however, that the
$\sigma_8$ we have adopted (based on WMAP-7) is somewhat lower than the favored
value from the first-year Planck results \citep{PlanckCosmo}, and even so the
number of massive failures remains high.

We have also checked for correlations with host mass, and find a weak positive
correlation, as expected from the scaling of the subhalo mass function.  The
scatter about the trend is very large, but an extrapolation of the fit suggests
that the MW mass must be below $\sim7\times10^{11}\msun$ to eliminate the
massive failures \citep[see also][]{MBK2012,Wang2012,Purcell2012}, which is in
conflict with large-scale dynamical mass estimates of the MW \citep[][and
references therein]{Marel2012,Boylan-Kolchin2013}.


\subsection{Massive failures in the Local Field}
\subsubsection{Counting discrepant field halos}
\label{sssec:fieldfails}

\begin{figure*}
\begin{minipage}[t]{\columnwidth}  
\centering
\vspace{0pt}
\includegraphics[width=\linewidth]{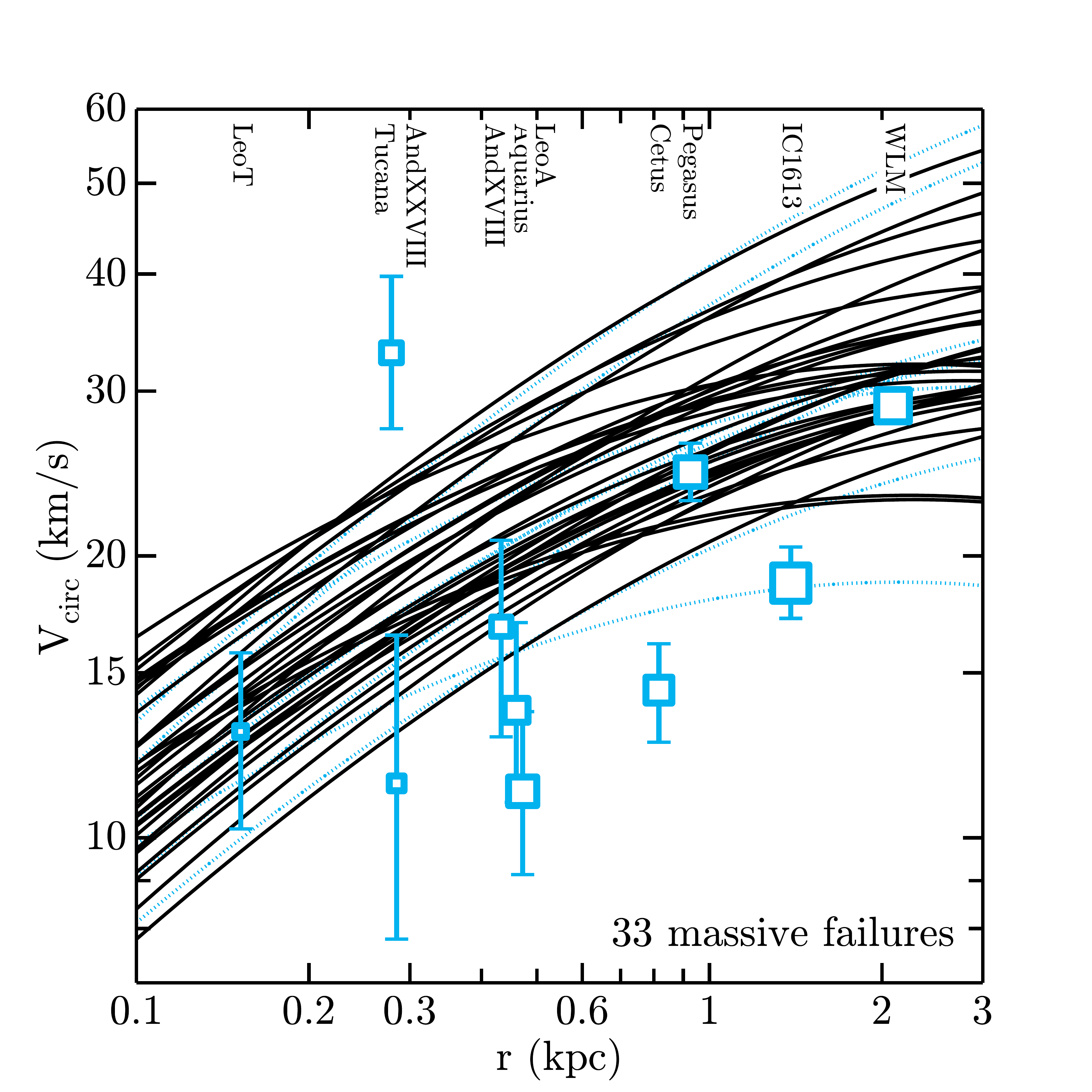}
\caption{Rotation curves ($\alpha=0.18$) for all resolved field halos 
in the LF around Zeus \& Hera with $\vpeak > 30~\kms$ 
(extrapolated from measured $\vmax$ and $\rmax$ values in the simulation).  Massive failures 
(unaccounted-for satellite halos that became large enough to from stars) 
are plotted as black lines; halos that are hosting one of the field dwarfs are 
indicated by light blue dotted lines.  As in Figure~\ref{fig:Rvircurves}, halos with 
$\vpeak < 30~\kms$ are not plotted~--~there are $254$ such resolved halos 
in the Local Field around Zeus \& Hera.  The light blue points indicate the kinematic 
constraints on the galaxies in the LF; their sizes are again proportional to the log of the 
stellar mass of each galaxy.  Many of the massive failures are denser than all the known 
field dwarfs except for Tucana.}
\label{fig:fieldcurves}
\end{minipage}
\hfill
\begin{minipage}[t]{\columnwidth}  
\centering
\vspace{0pt}
\includegraphics[width=\linewidth]{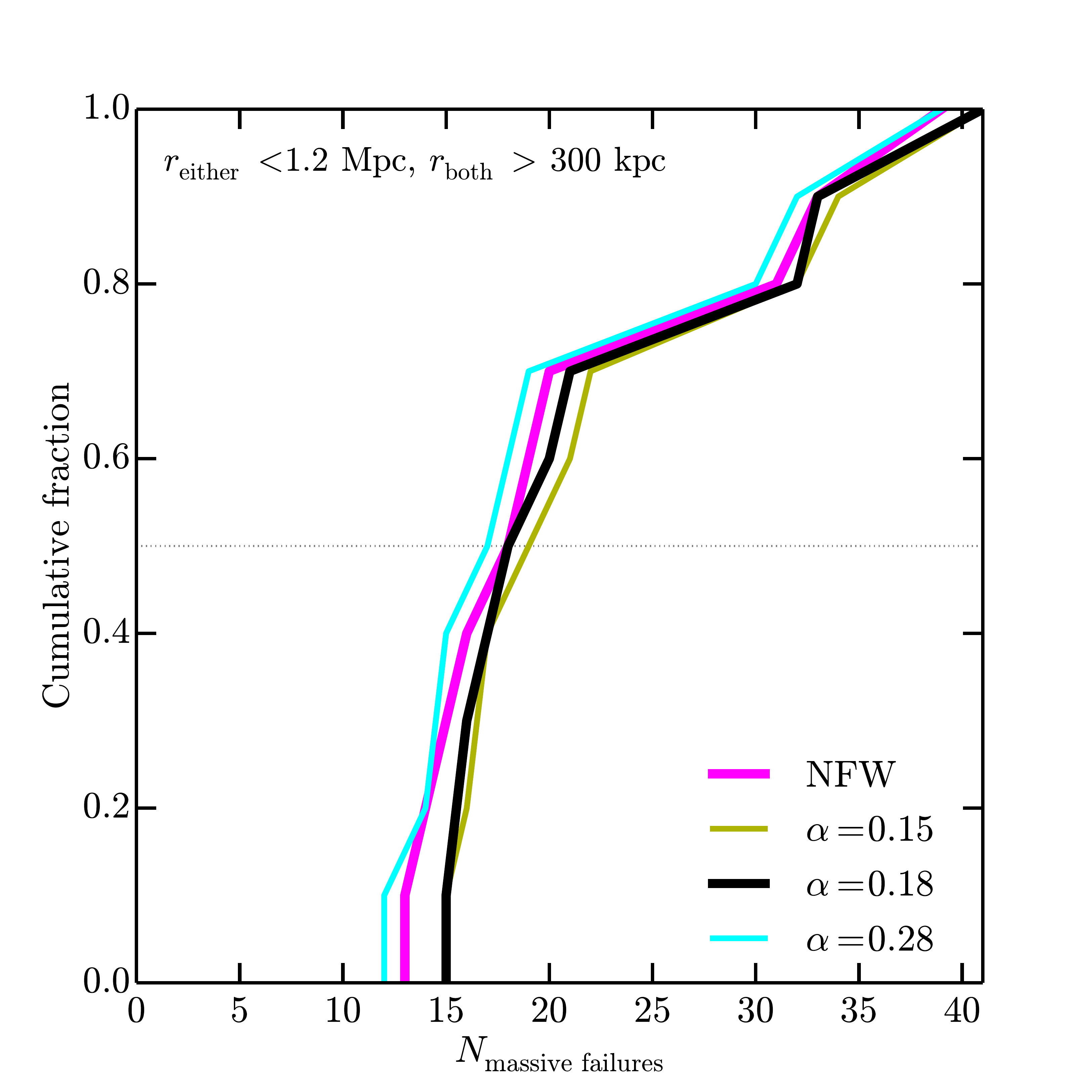}
\caption{The distribution of the number of massive failures in the fields
 surrounding the ten LG analogs in the ELVIS pairs without a third giant 
 nearby.  Plotted are the number of field halos with $\vpeak > 30~\kms$ 
 that do not have a corresponding bright galaxy in the field for the four profiles
 that we consider in this work; the colors are as in Figure~\ref{fig:cumhist310}.  
 The ELVIS simulations predict that there are $\sim18-20$ missing galaxies in the 
 Local Field, many of which should be denser than the majority of the known
 field dwarfs (i.e. comparable to Tucana and Leo~T).}
\label{fig:cumhistfield}
\end{minipage}
\end{figure*}

Now we extend our count of expected massive halos to the Local Field (LF) -- a
volume defined to be within 1.2 Mpc of either giant host, but excluding 300 kpc 
spherical regions around each in order to avoid satellites (and thus the potential 
for large tidal influences).  Figure~\ref{fig:fieldcurves} is analogous to
Figure~\ref{fig:Rvircurves}, in that it compares halos within the LF of the 
ELVIS pair Zeus \& Hera to observed galaxies within the same volume around 
the MW and M31.  In \citetalias{ELVIS}, we showed that the Zeus \& Hera pair 
provides a good match to the observed stellar mass function in the Local Group 
when abundance matching is applied (see Figure 9 of \citetalias{ELVIS}).  The 
open light blue data points plot constraints on $\vhalf$ at $\rhalf$ for the 
ten dark matter-dominated galaxies in the LF with measured line-of-sight stellar 
velocity dispersions, $\sstar$, again with sizes proportional to the log of their 
stellar masses.\footnote{See \S\ref{sssec:incompleteness} for a summary of the 
origin of the $\mstar$ estimates.}

There are four known galaxies that meet the distance cuts but that we 
exclude from our analysis:  NGC~6822, Sagittarius dIrr, Andromeda XVI, 
and Phoenix.  Of these four, all but NGC~6822 lack definitive mass measurements.
The galaxy NGC~6822 is baryon dominated \citep{Kirby2013} and 
we exclude it because determining its dark matter mass is difficult and
because  its host halo is likely to have undergone adiabatic 
contraction.    There have been no attempts to measure the stellar velocity
dispersion of the Sagittarius dIrr galaxy.   \citet{Letarte2009} established 
an upper limit of $\vhalf<17.3~\kms$ at $\rhalf = 0.18~\kpc$ for Andromeda~XVI, 
similar to the measurement for Leo~T in $(\vcirc,\,r)$ space.  In a conference
proceeding, \citet{Zaggia2011} published
 $(\vhalf,\rhalf) \approx (14~\kms,\,0.6~\kpc)$ for Phoenix, placing it 
between Aquarius and Cetus in Figure~\ref{fig:fieldcurves}, and therefore
among the lower density dwarfs.  Therefore, 
our massive failure counts may be high by $3$ (before accounting for incompleteness, 
which we discuss further in \S\ref{sssec:incompleteness}).

For the seven galaxies that are purely dispersion supported, we calculate 
$\vhalf$ from $\sstar$ via the \citet{Wolf2010} formula.  Velocity dispersions 
for the two Andromeda dwarfs with constraints on $\sstar$ that meet the 
distance cuts are from \citet{Collins2013}.  Measurements for the field dwarfs 
are from \citet{Kirby2013} where available; the constraints on Leo T and Tucana 
are from \citet{Simon2007} and \citet{Fraternali2009}, respectively.  Three of 
the field dwarfs~--~WLM, Pegasus, and Tucana~--~also display evidence of rotation
support, and are therefore not well described by the \citet{Wolf2010} methodology.  
We use the result from \citet{Leaman2012} for WLM, who calculated the mass 
within $\rhalf$ with a detailed dynamical model.  For the latter two, we follow 
\citet{Weiner2006} in replacing $\sigma_{\star}^2$ with 
$\sstar^2+\frac{1}{2}(v\sin i)^2$ when calculating $\vhalf$, where 
$v\sin i$ is the projected rotation velocity  \citep[see also \S5.2 of][]{Kirby2013}.

\begin{figure*}
\begin{minipage}[t]{\columnwidth} 
\vspace{0pt}
\centering
\includegraphics[width=\linewidth]{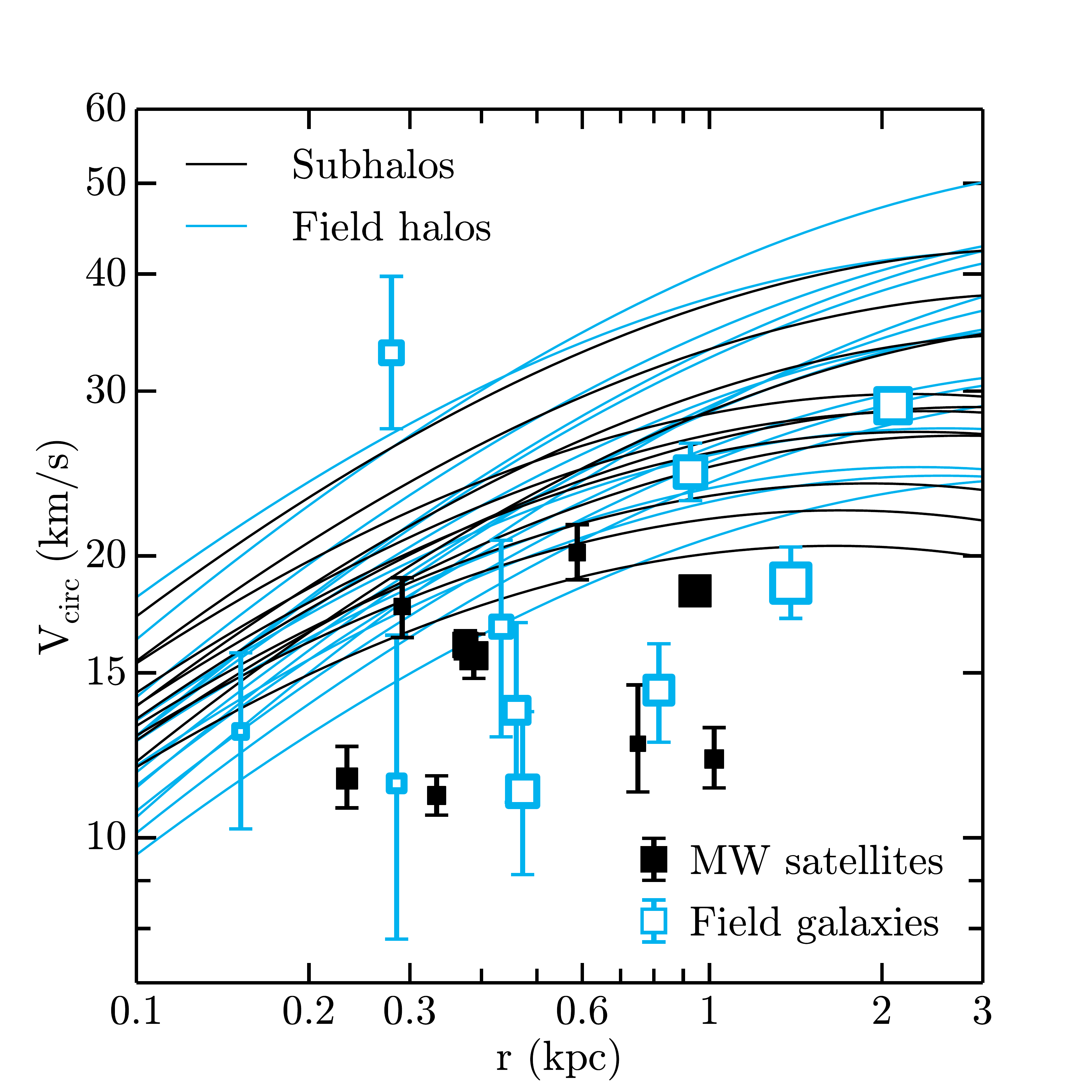}
\caption{Plotted are the rotation curves for all halos identified
as massive failures around Douglas, both within $300~\kpc$ (black lines)
and in the Local Field surrounding it (light blue lines), along with
constraints on the dwarf galaxies in each region (black squares denote
MW satellites and open light blue squares indicate field galaxies~--~sizes 
are again proportional to $\mstar$); i.e. combining Figure~\ref{fig:Rvircurves} 
with a plot equivalent to Figure~\ref{fig:fieldcurves}.  Explicitly excluded are
halos with $\vpeak < 30~\kms$; also not plotted are the halos selected to 
host a galaxy.}
\label{fig:allcurves}
\end{minipage}
\hfill
\begin{minipage}[t]{\columnwidth}
\vspace{0pt}
\centering
\includegraphics[width=\linewidth]{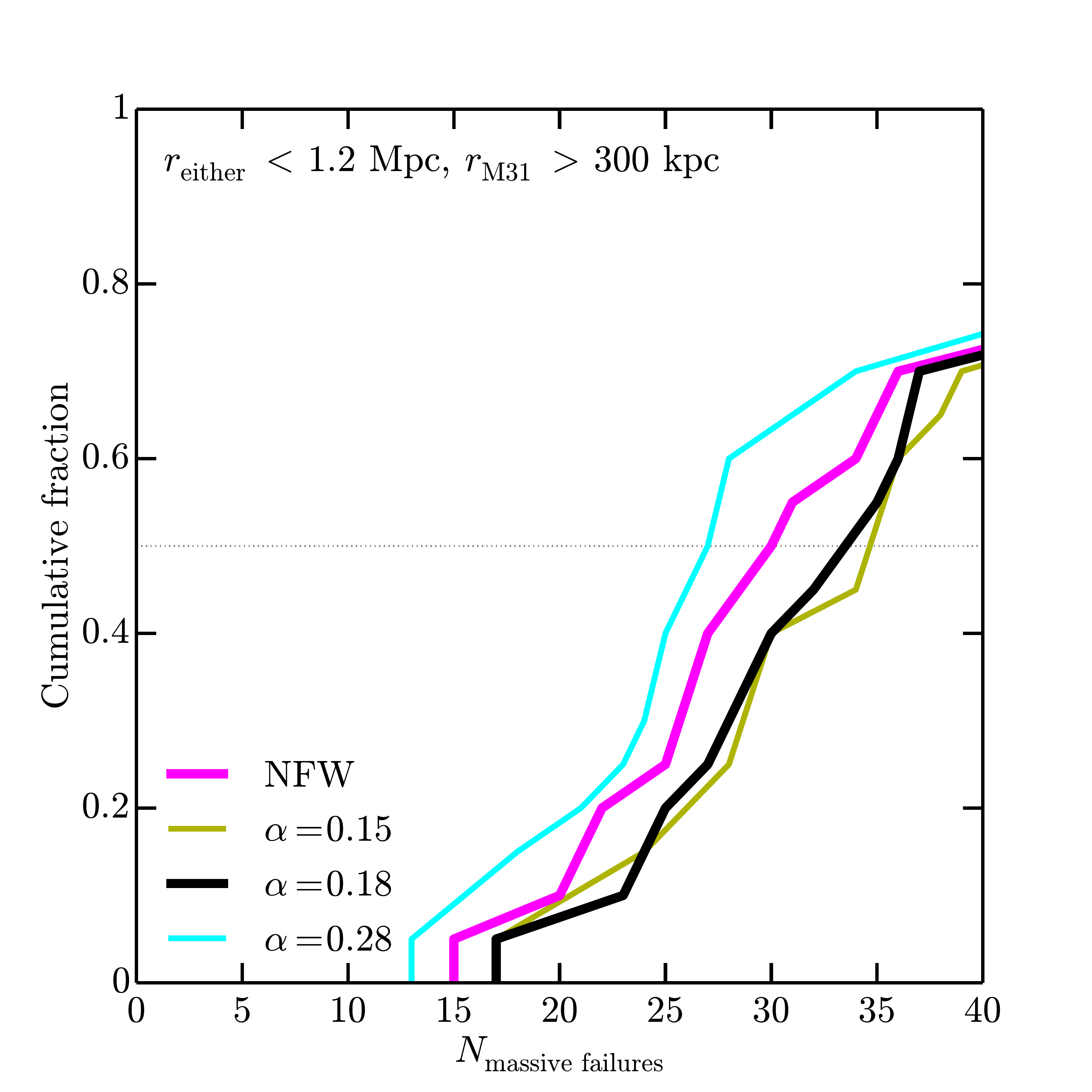}
\caption{The distribution of the number of massive failures in each of twenty
paired halos plus the field around them, i.e. combining results from 
Figures~\ref{fig:cumhist310}~and~\ref{fig:cumhistfield} but excluding 
failures that are within $300~\kpc$ of the M31 analog.  Colors are as in 
Figures~\ref{fig:cumhist310}~and~\ref{fig:cumhistfield}.  The exact
number of massive failures depends on the specific density profile, but
the conclusion that there are many missing large, dense halos in the Local
Field is robust:  each system has at least $14$ massive failures, with a median 
between $\sim26-34$.}
\label{fig:cumhisttot}
\end{minipage}
\end{figure*}

The lines in Figure~\ref{fig:fieldcurves} plot the extrapolated rotation curves of the 
resolved dwarf halos with $\vpeak > 30~\kms$ around Zeus \& Hera, again 
assuming an Einasto profile with $\alpha = 0.18$.  That the lower-right section 
of the plot is empty is typical of the ELVIS fields~--~only $\sim$10-25\%    
of the field halos that meet the ``massive" cut ($\vpeak > 30~\kms$) have 
been sufficiently stripped to have $\vmax < 25~\kms$.    Blue dotted lines indicate 
individual halos that are consistent with observed dwarfs; we do not count 
these systems as massive failures.

The black lines in Figure~\ref{fig:fieldcurves} indicates the massive failures
in the Local Field.  Due strictly to the published mass for Tucana, which is above 
\emph{every} halo in the sample for $\alpha=0.18$; there are no strong
massive failures in the LFs around the ELVIS hosts.\footnote{The field around
Scylla \& Charybdis contains two halos with circular velocities that marginally 
exceed that of Tucana at $\rhalf$ if $\alpha=0.15$, but they agree within $1\sigma$.}
However, the systematic over-abundance of large halos remains:  though Tucana
eliminates any strong massive failures in the LF, the median number of halos
per field that are consistent \emph{only} with Tucana, i.e. the number of halos that
would be identified as strong massive failures if Tucana did not exist, is $7.5$, again
assuming $\alpha = 0.18$.  We will further show below that, if abundance matching 
holds at these masses, most of these galaxies should be bright ($\mstar>10^6\msun$).
Moreover, the lack of environmental stripping at larger radii leaves the vast majority
of these objects with $\vmax>30~\kms$ today.

The distribution of the number of massive failures in the Local Field is plotted in 
Figure~\ref{fig:cumhistfield}.  The number of halos that are naively expected 
to host luminous galaxies ($\vpeak > 30~\kms$) exceeds the number of 
known dwarfs by a factor $\gtrsim2$ in every case~--~no system has fewer than thirteen 
massive failures, even for $\alpha = 0.28$.  Importantly, the exact number is insensitive 
to the assumed profile, with the minimum count of massive failures varying only by 
$\pm 3$ among the pairs studied here.   In a relative sense, the LF massive failure 
counts are even more robust than the counts within 300 kpc.  
The minimum number of massive failures in the LF varies from $12-15$ (depending on assumed 
profile shape) and the median number varies from $16-18$.\footnote{Unlike the situation within 
$300~\kpc$, the missing halos are not explained by cored profiles:  due to the 
relatively large half-light radii of WLM and IC~1613, there are at least eleven 
massive failures in each LF, even assuming $\alpha = 1$.}

Of course, the count given in Figure~\ref{fig:cumhistfield} ignores massive failures
within the virial radii of either M31 or the MW.  In order to give a more complete 
picture of TBTF problem throughout the Local Group, Figure~\ref{fig:allcurves} 
plots the rotation curves of all the massive failures near Douglas (excluding 
only those within $300~\kpc$ of its M31 analog, Lincoln); i.e. it combines 
Figure~\ref{fig:Rvircurves} with a plot equivalent to Figure~\ref{fig:fieldcurves}.  
Plotted as black lines are massive failures within $300~\kpc$; the light blue lines 
plot massive failures in the LF.  The black and light blue points again plot 
constraints on the MW satellites and galaxies in the LF, respectively.  Halos 
selected to host those galaxies are not plotted.   We have not included a 
comparison of the full Local Group including M31 satellites because, as explained
above, M31 contains several baryon-dominated satellites, making the accounting 
more complicated.  A more in-depth analysis of the M31 system is given in 
\citet{Tollerud2014}.

Figure~\ref{fig:cumhisttot} provides an overview of the TBTF problem in the LG.
As before, we combined the results of Figures~\ref{fig:cumhist310}~and~\ref{fig:cumhistfield},
adding together the counts within 
$300~\kpc$ and the Local Field for each MW analog, excluding the $300$ kpc 
volume around the M31 analog.  The distribution is therefore based on twenty virial 
volumes combined with ten LF analogs; none of these combinations contain fewer 
than thirteen massive failures.  We find typically $\sim26-34$ massive 
failures in the Local Volume, even excluding halos and galaxies within $300~\kpc$ of M31.
We find no trend between the number of massive failures within $300~\kpc$
of a host and the number within the LF surrounding it.  

Tides from disk interactions and ram pressure stripping are baryonic process that have been invoked to lower the density of
massive failure halos beyond what is predicted in dissipationless simulations
 \citep{Zolotov2012,Arraki2012,BrooksZolotov2012,Brooks2013}. 
However, in the Local Field, 
particularly more than $\sim500~\kpc$ from the nearest giant where the backsplash 
fraction is below $50\%$ \citepalias{ELVIS}, central halo densities should remain
largely unaffected by tidal and ram pressure stripping.  Moreover, Tucana, which shows evidence
of having interacted with the MW \citep{Teyssier2012}, is the most dense galaxy
in the field, calling into question proposed environmental mechanisms.
Galaxies large enough to have affected their density profiles 
via supernovae feedback may be lurking unseen on the outer edge of the LF, but 
no galaxies brighter than $10^7\lsun$ have been discovered in the LF within 
the past fifty-five years \citep[Pegasus dIrr;][]{Holmberg1958}.

\begin{figure*}
\centering
\begin{minipage}[t]{\columnwidth}
\vspace{0pt}
\centering
\includegraphics[width=\linewidth]{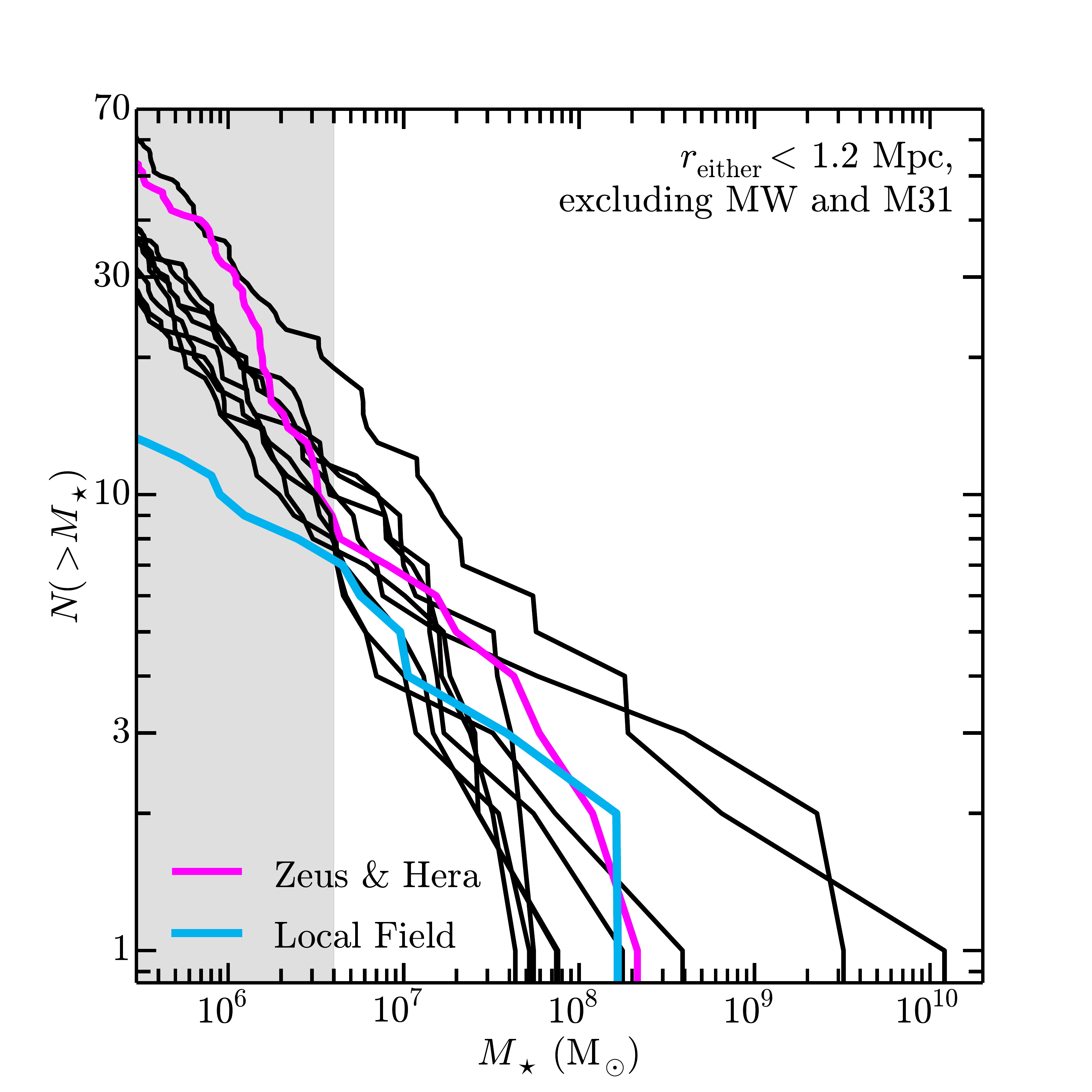}
\caption{The stellar mass function observed in the Local Field (light blue)
along with the stellar mass functions in the fields surrounding the ELVIS pairs, 
assuming the AM relation presented in \citetalias{ELVIS}.  The shaded region 
indicates stellar masses where the current census of galaxies lies below 
that of all the ELVIS pairs, $\mstar < 4\times10^6~\msun$; at this mass, 
however, the count of known field galaxies nearly matches that around Zeus \& Hera 
(highlighted in magenta), the LF shown in Figure~\ref{fig:fieldcurves}.}
\label{fig:mstarfunc}
\end{minipage}
\hfill
\begin{minipage}[t]{\columnwidth}
\vspace{0pt}
\centering
\includegraphics[width=\linewidth]{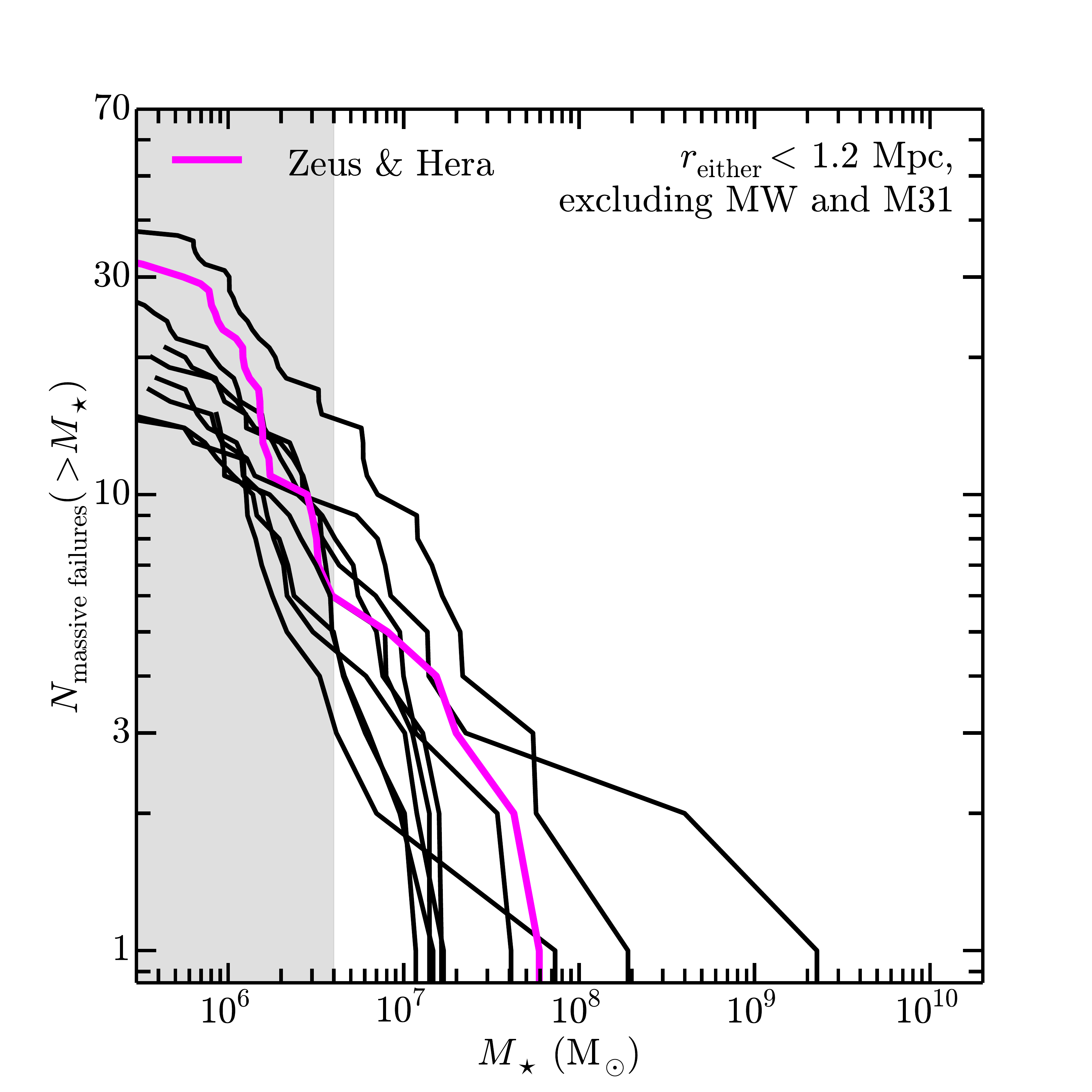}
\caption{The stellar mass function, again from abundance matching, of the 
halos identified as massive failures in the LF; i.e. the black lines in 
Figure~\ref{fig:fieldcurves} counted in Figure~\ref{fig:cumhistfield}.
The magenta line again highlights the LF around Zeus \& Hera.  Even selecting 
those halos with the highest possible $\mstar$ to host the known dwarfs, 
the massive failures stellar mass functions are largely unchanged at the high 
mass end from Figure~\ref{fig:mstarfunc}.  Therefore, although the number
count agree from $\mstar \gtrsim 10^{6.5}\msun$, only lower mass field halos 
are kinematically compatible with the known LF galaxies.}
\label{fig:mstarfails}
\end{minipage}
\end{figure*}

\subsubsection{Missing galaxies in the Local Field?}
\label{sssec:incompleteness}

In this Section, we present the stellar masses of those halos identified as
massive failures, from abundance matching, and investigate whether the they can
be explained as unidentified dwarf galaxies in the LF.  Though no galaxies have
been discovered within the distance cut since the discovery of Andromeda XXVIII
\citep{Slater2011}, the recent discovery of Leo P
\citep{Giovanelli2013,Rhode2013} at a distance of $\sim1.5~\mpc$ from the MW
suggests that there may be new galaxies in the Local Volume that will be
identified via HI observations or upcoming deep stellar surveys.

We begin by plotting the predicted stellar mass functions implied by our favored
AM extrapolation from \citetalias{ELVIS}, along with the observed stellar mass
function of galaxies that meet the same radial cuts in the LG (in blue) in
Figure~\ref{fig:mstarfunc}.\footnote{We emphasize that the stellar mass range
  shown is large enough that an AM-inspired power-law relationship between
  $\mstar$ and $\mvir$ is well-motivated. Specifically, this is above the mass
  regime ($\mstar < 10^6~\msun$, $\vpeak < 30~\kms$) where processes like
  reionization might act to ``bend" the relation \citep{Sawala2014,ELVIS},
  possibly suppressing the count of faint galaxies in the Local Group.}  Stellar
masses are from \citet{Woo2008} where available and are otherwise taken from the
data cataloged in \citet{McConnachie2012}, assuming $\mstar/L = 2$.  We
emphasize that the adopted AM relation does well in reproducing the observed
stellar mass function above stellar masses $\mstar = 4\times10^6\msun$.  The
shaded region below this point draws attention to the region where the known
census of galaxies lies below that predicted.  Above this mass, however, the
galaxy count around Zeus \& Hera, the pair plotted in
Figure~\ref{fig:fieldcurves} and highlighted in magenta in
Figure~\ref{fig:mstarfunc}, nearly matches that observed in the LF.

While a simple extrapolation of abundance matching creates a stellar mass
function that agrees well with galaxy counts, {\em it does so by matching
  galaxies with halos that are too dense to reproduce the observed kinematics of
  those same galaxies} \citep[see also][]{MBK2012}.  Specifically, it is
difficult to match both the observed luminosity function and the observed
densities of galaxies at the same time.  The magnitude of the problem is
demonstrated explicitly in Figure~\ref{fig:mstarfails}, which plots the stellar
mass function of only the halos identified as massive failures (i.e. the stellar
masses associated with the black lines in Figure~\ref{fig:fieldcurves},
specifically with $\alpha = 0.18$.)  This is the subset of the stellar mass
function\footnote{When selecting hosts for each galaxy, the candidate halos were
  sorted by $\mstar$~--~that is, the halos plotted in
  Figure~\ref{fig:mstarfails} are selected to have the {\em smallest} possible
  stellar masses.  Nonetheless, the high mass end is largely unchanged from
  Figure~\ref{fig:mstarfunc}, clearly showing that many of the massive failures
  are among the highest mass halos in the field and would naively be expected to
  host {\em bright} galaxies.}  shown in Figure~\ref{fig:mstarfunc} that
includes only $\vpeak > 30~\kms$ halos that remain dense today ($\vmax \gtrsim
25\, \kms$) and that are unaccounted for by any known galaxy.  The takeaway
point from Figure \ref{fig:mstarfails} is this: the TBTF halos should naively be
hosting fairly bright galaxies, many of which should be more massive than
$\mstar \simeq 5 \times 10^6 \msun$.

As we show in the next section, based on the densities measured, the stellar
mass of a galaxy does not seem to scale at all with the maximum circular
velocity of the dark matter halo that it resides in.  In the absence of baryonic
processes that strongly affect halo densities, it is hard to understand how the
relation could be as stochastic as it appears to be.

\subsection{The $\mathbf{\vmax}$-$\mathbf{\mstar}$ relation in the Local Field}
\label{sssec:mstarvmax}
As the previous sections showed, it is likely that either there are roughly $15$ 
dense galaxies living in high $\vmax$ halos in the Local Field that have yet to be 
discovered, or that the densities of $\mstar \sim 10^{6.5} \msun$ {\em field} galaxies 
are much less dense than expected from straightforward $\Lambda$CDM predictions.

In this subsection, we make this point explicitly by working out the inferred relationship 
between galaxy stellar mass and dark matter halo mass under the assumption that LF 
halos are unaffected by baryonic processes, and then compare that relationship to that 
expected from AM in the same volume.

\begin{figure}
\centering
\includegraphics[width=0.49\textwidth]{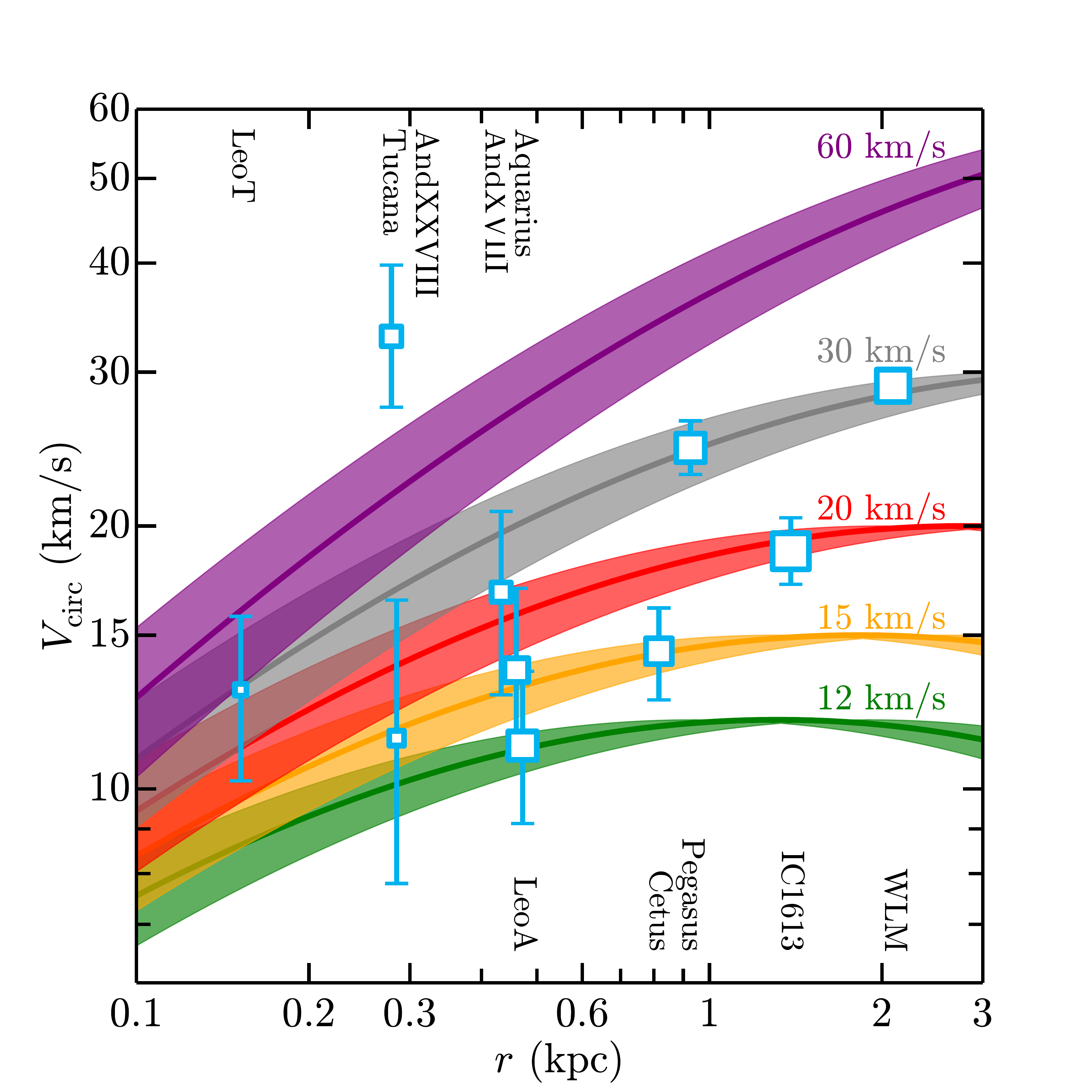}
\caption{Typical rotation curves of halos in the Local Field for  $\alpha = 0.18$, 
from the relations in Figure~\ref{fig:rvfield}.  Also plotted as open light 
blue points are the ten galaxies in the LF used in \S~\ref{sssec:fieldfails} as in 
Figures~\ref{fig:fieldcurves}~and~\ref{fig:allcurves}; the points are again sized
according to their stellar masses.  The stellar masses of the halos do not appear to 
scale with $\vmax$, assuming a universal density profile.}
\label{fig:typcurves}
\end{figure}

Our approach is demonstrated in Figure~\ref{fig:typcurves}, where the shaded
bands show typical rotation curves for halos of various $\vmax$ values.  The
width of the bands correspond to the $1\sigma$ scatter $\rmax$ at fixed
$\vmax$ given in Equation~\ref{eqn:310fit} and Table~\ref{tab:fieldnorms},
assuming Einasto profiles with $\alpha=0.18$.   The points correspond to dwarfs
and are identical 
to those in Figure~\ref{fig:fieldcurves} with sizes that are again proportional to 
their stellar masses.  Note that the least luminous dwarf (Leo T) appears to
reside in a fairly massive ($\vmax \simeq 30 \kms$) halo, while the galaxy
IC1613, which is $\sim 1000$ times more luminous, appears to reside in a halo that
is less massive ($\vmax \simeq 20 \kms$).  Given the large errors in Leo T's mass,
the inferred halo sizes could be equal, but if there is any positive correlation between
halo $\vmax$ and stellar mass, it must be extremely weak.

\begin{figure}
\centering
\includegraphics[width=\linewidth]{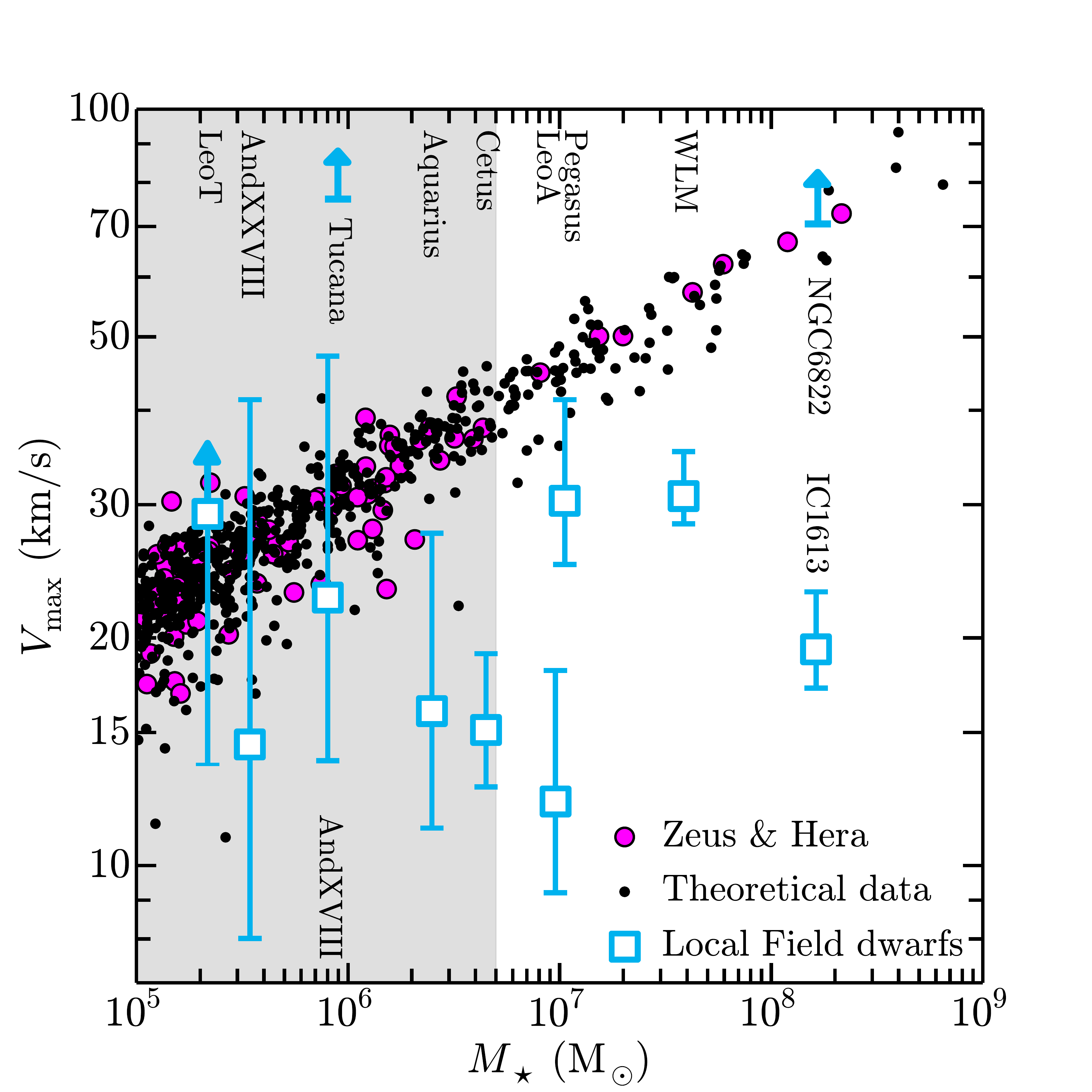}
\caption{A comparison between the best-fit values of $\vmax$ 
(assuming $\alpha = 0.18$) of the Local Field dwarfs to the 
stellar masses implied by the preferred AM relation in \citetalias{ELVIS}.  
As expected, the latter follow a power law; the scatter is due to the 
scatter between $\vmax$ and $\mpeak$, upon which stellar masses 
are based.  The former, however, appear to follow an extremely weak 
trend, indicating that stellar mass may not scale with $\vmax$ at these 
low luminosities.  Halos near Zeus \& Hera are highlighted in magenta; 
the shaded region is the same as that in Figure~\ref{fig:mstarfunc}.  
Due to the scaling of $\rmax$ with $\vmax$, the measurement for Tucana
is incompatible with the median relation; the $1\sigma$ bound is indicated
by the arrow.  Similarly, Leo~T is unconstrained at the upper-end.  The $1\sigma$
lower limit for NGC~6822 is also indicated, though it is baryon dominated 
and unlikely to be well described by an Einasto profile. }
\label{fig:mstarvmax}
\end{figure}

How does the implied relation compare to that expected from abundance matching?
In Figure~\ref{fig:mstarvmax} we quantify the inferred relation, using the observational
errors on dwarf masses together with the scatter in $\rmax$ at fixed $\vmax$ measured 
for LF halos in the ELVIS suite.  Specifically, we plot the inferred $\vmax$ for each LF galaxy
as a function of $\mstar$ as open light blue points.  Error bars are $1\sigma$.  
Due to its small half-light radius, Leo~T may be hosted by any halo with $\vmax\gtrsim14~\kms$ at
the $1\sigma$ level, though the median relation predicts that it is hosted by
a halo with $\vmax=29~\kms$.  The upward arrows indicate the lower 
limits for Tucana and NGC~6822.  Assuming the median relation between 
$\rmax$ and $\vmax$, Tucana is incompatible with an Einasto profile with 
$\alpha = 0.18$ for all values of $\vmax$, though it may be hosted by a 
halo that is only a $1\sigma$ outlier.  NGC~6822, as mentioned above, is 
dominated by baryonic mass within $\rhalf$ and is therefore unlikely to follow 
either an Einasto or NFW profile.

The circles in Figure~\ref{fig:mstarvmax} 
indicate theoretical expectations from the AM relation in \citetalias{ELVIS}, the same
relation that produces the {\em observationally-consistent} stellar mass function shown in Figure~\ref{fig:mstarfunc}.
The magenta circles highlight  those halos around Zeus \& Hera -- the same halos that
have a stellar mass function that masses the Local Group well in Figure~\ref{fig:mstarfunc}. 

Assuming that galaxies in the Local Field have density profiles of the kind
predicted in our dissipationless simulations, any relation between $\vmax$ 
and $\mstar$ for galaxies in the LF must be very weak (also see \citealt{Strigari2008}
and \citealt{MBK2012}, who found similar results for MW satellites).
This may suggest that the scaling between halo mass and stellar mass breaks 
down for small $\mstar \lesssim 10^8 \msun$, but if the underlying relation 
followed something close to $\mstar \sim \vmax^0$ over the mass range shown 
(and with a scatter similar to that shown in the data plotted) then this would 
drastically over-predicted the number of $\mstar \sim 10^{6.5} \msun$ 
galaxies in the Local Group.

Another option is that the shape of the density profiles of the halos hosting LF
galaxies vary strongly from system to system.  Because these galaxies exist in the
field, tidal interactions and ram pressure stripping will not strongly affect their 
dark matter halos.  Moreover, unless these galaxies formed with top-heavy initial 
mass functions or live in much smaller halos than abundance matching suggests, 
the energy available from supernovae is likely below that required to alter their 
density profiles significantly \citep{Penarrubia2012,Garrison-Kimmel2013}.

We caution that the error bars in Figure~\ref{fig:mstarvmax} account only for 
the observational errors on $\vhalf$ and for the scatter in the $\rmax-\vmax$ 
relationship; that is, we are requiring that all galaxies reside in halos with
identical density profile shapes.  Additionally, we impose no sampling prior based on the
predicted number of halos of a given $\vmax$, which would serve to shrink the error bars
in Figure~\ref{fig:mstarvmax} and systematically push some of the inferred $\vmax$ 
values lower \citep{Martinez2013}.  A more detailed analysis should be performed, 
but we leave that effort for future work.

\section{Conclusions}
\label{sec:conclusions}
In this paper, we have analyzed the structural properties of the small halos
in the ELVIS Suite~--~both those within the virialized volumes of the two
giant halos and those in the fields surrounding them.  Our results indicate
that the Too Big to Fail problem, the discrepancy in central masses between
the large subhalos of simulated MWs and the dSphs surrounding the MW,
is an issue not only within $300~\kpc$, where environmental physics may be
able to resolve the disagreement, but also in the Local Field, where such
effects should be small.  Specifically, we find that

\begin{itemize}
\item For NFW-like density profiles, nearly all of the ELVIS hosts contain at
  least one ``strong massive failure"~--~satellite halos that are too dense to
  host any of the classical dSphs.  The median number of strong massive failures
  per host is highly dependent on the assumed density profile, varying between 2
  and 10, and would change dramatically if a dwarf much denser than Draco is
  discovered.
\item The number of ``massive failures," $\vpeak > 30~\kms$ halos that remain
  dense at z = 0 and cannot be accounted for with the known census of dSphs, is
  much less dependent on the assumed profile.  All of the ELVIS hosts contain at
  least one massive failure for the profiles considered in the work, with a median
  varying between 8.5 and 13.  Unlike the count of strong massive failures, a newly
  discovered high-density dwarf would only alter these numbers by one.
\item Though there are typically no strong massive failures in the Local Field
  (i.e.  more than $300~\kpc$ from both giants in the LG), the overall
  discrepancy between known galaxies that appear to live in dense (typically
  high mass) halos and the number of these halos predicted is even stronger.
  Most simulated LFs contain $\gtrsim15$ more of these dense halos than can be
  accounted for observationally.
\item If the discrepancy is to be resolved by discovering new galaxies, and
  \emph{if} the stellar mass of a galaxy scales in a reasonable way with
  $\vmax$, then the abundance matching technique predicts that there should be
  $\sim2-10$ undiscovered galaxies with $\mstar>10^7\msun$ within the LF, though
  there have been none found since 1958.  However, perhaps more puzzlingly, the
  stellar masses of the known field galaxies do not appear to correlate with the
  apparent $\vmax$ of their host halos, as estimated from $\vhalf$, suggesting
  either that the density profiles of the dwarfs vary strongly or that the
  scaling of $\mstar$ with $\vmax$ breaks down at low luminosities.
\end{itemize}

The results presented in this work do not necessarily indicate the need to move
beyond the standard \lcdm\ model with collisionless dark matter.  They can
largely be viewed as predictions for results from future surveys, such as LSST
and DES. However, if these missing dense galaxies are not discovered as we probe
the nearby Universe to an increasing depth, these large dark matter halos must
somehow be explained.

\vskip1cm

\noindent {\bf{Acknowledgments}} \\
The authors thank Manoj Kaplinghat, Anna Nierenberg, Mike Cooper, Erik Tollerud,
Arianna Di Cintio, Shunsaku Horiuchi, and Jose O{\~n}orbe for helpful discussions.  

Support for this work was provided by NASA through a
\textit{Hubble Space Telescope} theory grant (program AR-12836) from the Space
Telescope Science Institute (STScI), which is operated by the Association of
Universities for Research in Astronomy (AURA), Inc., under NASA contract
NAS5-26555. This work was also supported by a matching equipment grant from
UC-HiPACC, a multicampus research program funded by the University of California
Office of Research.

We also acknowledge the computational support of the NASA Advanced 
Supercomputing Division and the NASA Center for Climate Simulation, 
upon whose \textit{Pleiades} and \textit{Discover} systems the ELVIS 
simulations were run, and the \textit{Greenplanet} cluster at UCI, upon 
which much of the secondary analysis was performed.

\bibliographystyle{mn2e}
\bibliography{elvis_ii.bib}

\begin{thebibliography}{123}
\expandafter\ifx\csname natexlab\endcsname\relax\def\natexlab#1{#1}\fi

\bibitem[{{Agnello} \& {Evans}(2012)}]{Agnello2012}
{Agnello} A., {Evans} N.~W., 2012, \apjl, 754, L39

\bibitem[{{Amorisco} {et~al}\mbox{.}(2013{\natexlab{a}}){Amorisco}, {Agnello},
  \& {Evans}}]{Amorisco2013}
{Amorisco} N.~C., {Agnello} A., {Evans} N.~W., 2013{\natexlab{a}}, \mnras, 429,
  L89

\bibitem[{{Amorisco} {et~al}\mbox{.}(2013{\natexlab{b}}){Amorisco}, {Zavala},
  \& {de Boer}}]{Amorisco2013feedback}
{Amorisco} N.~C., {Zavala} J., {de Boer} T.~J.~L., 2013{\natexlab{b}},
  {arXiv:1309.5958 [astro-ph]}

\bibitem[{{Anderhalden} {et~al}\mbox{.}(2013){Anderhalden}, {Schneider},
  {Macci{\`o}}, {Diemand}, \& {Bertone}}]{Anderhalden2013}
{Anderhalden} D., {Schneider} A., {Macci{\`o}} A.~V., {Diemand} J., {Bertone}
  G., 2013, \jcap, 3, 14

\bibitem[{{Arraki} {et~al}\mbox{.}(2013){Arraki}, {Klypin}, {More}, \&
  {Trujillo-Gomez}}]{Arraki2012}
{Arraki} K.~S., {Klypin} A., {More} S., {Trujillo-Gomez} S., 2013, \mnras

\bibitem[{{Babul} \& {Rees}(1992)}]{Babul1992}
{Babul} A., {Rees} M.~J., 1992, \mnras, 255, 346

\bibitem[{{Behroozi} {et~al}\mbox{.}(2013c){Behroozi}, {Wechsler}, \&
  {Conroy}}]{BehrooziAM}
{Behroozi} P.~S., {Wechsler} R.~H., {Conroy} C., 2013c, \apj, 770, 57

\bibitem[{{Behroozi} {et~al}\mbox{.}(2013a){Behroozi}, {Wechsler}, \&
  {Wu}}]{rockstar}
{Behroozi} P.~S., {Wechsler} R.~H., {Wu} H.-Y., 2013a, \apj, 762, 109

\bibitem[{{Behroozi} {et~al}\mbox{.}(2013b){Behroozi}, {Wechsler}, {Wu},
  {Busha}, {Klypin}, \& {Primack}}]{Behroozi2013b}
{Behroozi} P.~S., {Wechsler} R.~H., {Wu} H.-Y., {Busha} M.~T., {Klypin} A.~A.,
  {Primack} J.~R., 2013b, \apj, 763, 18

\bibitem[{{Belokurov} {et~al}\mbox{.}(2007){Belokurov}, {Zucker}, {Evans},
  {Kleyna}, {Koposov}, {Hodgkin}, {Irwin}, {Gilmore}, {Wilkinson}, {Fellhauer},
  {Bramich}, {Hewett}, {Vidrih}, {De Jong}, {Smith}, {Rix}, {Bell}, {Wyse},
  {Newberg}, {Mayeur}, {Yanny}, {Rockosi}, {Gnedin}, {Schneider}, {Beers},
  {Barentine}, {Brewington}, {Brinkmann}, {Harvanek}, {Kleinman}, {Krzesinski},
  {Long}, {Nitta}, \& {Snedden}}]{Belokurov2007}
{Belokurov} V. {et~al.}, 2007, \apj, 654, 897

\bibitem[{{Belokurov} {et~al}\mbox{.}(2006){Belokurov}, {Zucker}, {Evans},
  {Wilkinson}, {Irwin}, {Hodgkin}, {Bramich}, {Irwin}, {Gilmore}, {Willman},
  {Vidrih}, {Newberg}, {Wyse}, {Fellhauer}, {Hewett}, {Cole}, {Bell}, {Beers},
  {Rockosi}, {Yanny}, {Grebel}, {Schneider}, {Lupton}, {Barentine},
  {Brewington}, {Brinkmann}, {Harvanek}, {Kleinman}, {Krzesinski}, {Long},
  {Nitta}, {Smith}, \& {Snedden}}]{Belokurov2006}
{Belokurov} V. {et~al.}, 2006, \apjl, 647, L111

\bibitem[{{Boylan-Kolchin} {et~al}\mbox{.}(2011){Boylan-Kolchin}, {Bullock}, \&
  {Kaplinghat}}]{MBK2011}
{Boylan-Kolchin} M., {Bullock} J.~S., {Kaplinghat} M., 2011, \mnras, 415, L40

\bibitem[{{Boylan-Kolchin} {et~al}\mbox{.}(2012){Boylan-Kolchin}, {Bullock}, \&
  {Kaplinghat}}]{MBK2012}
{Boylan-Kolchin} M., {Bullock} J.~S., {Kaplinghat} M., 2012, \mnras, 422, 1203

\bibitem[{{Boylan-Kolchin} {et~al}\mbox{.}(2013){Boylan-Kolchin}, {Bullock},
  {Sohn}, {Besla}, \& {van der Marel}}]{Boylan-Kolchin2013}
{Boylan-Kolchin} M., {Bullock} J.~S., {Sohn} S.~T., {Besla} G., {van der Marel}
  R.~P., 2013, \apj, 768, 140

\bibitem[{{Boylan-Kolchin} {et~al}\mbox{.}(2010){Boylan-Kolchin}, {Springel},
  {White}, \& {Jenkins}}]{Boylan-Kolchin2010}
{Boylan-Kolchin} M., {Springel} V., {White} S.~D.~M., {Jenkins} A., 2010,
  \mnras, 406, 896

\bibitem[{{Boylan-Kolchin} {et~al}\mbox{.}(2009){Boylan-Kolchin}, {Springel},
  {White}, {Jenkins}, \& {Lemson}}]{Boylan-Kolchin2009}
{Boylan-Kolchin} M., {Springel} V., {White} S.~D.~M., {Jenkins} A., {Lemson}
  G., 2009, \mnras, 398, 1150

\bibitem[{{Breddels} \& {Helmi}(2013)}]{Breddels2013}
{Breddels} M.~A., {Helmi} A., 2013, \aap, 558, A35

\bibitem[{{Brook} {et~al}\mbox{.}(2013){Brook}, {Di Cintio}, {Knebe},
  {Gottl{\"o}ber}, {Hoffman}, {Yepes}, \& {Garrison-Kimmel}}]{Brook2013}
{Brook} C.~B., {Di Cintio} A., {Knebe} A., {Gottl{\"o}ber} S., {Hoffman} Y.,
  {Yepes} G., {Garrison-Kimmel} S., 2013, {arXiv:1311.5492 [astro-ph]}

\bibitem[{{Brooks} {et~al}\mbox{.}(2013){Brooks}, {Kuhlen}, {Zolotov}, \&
  {Hooper}}]{Brooks2013}
{Brooks} A.~M., {Kuhlen} M., {Zolotov} A., {Hooper} D., 2013, \apj, 765, 22

\bibitem[{{Brooks} \& {Zolotov}(2012)}]{BrooksZolotov2012}
{Brooks} A.~M., {Zolotov} A., 2012, {arXiv:1207.2468 [astro-ph]}

\bibitem[{{Bullock}(2010)}]{Bullock2010}
{Bullock} J.~S., 2010, {arXiv:1009.4505 [astro-ph]}

\bibitem[{{Bullock} {et~al}\mbox{.}(2000){Bullock}, {Kravtsov}, \&
  {Weinberg}}]{Bullock2000}
{Bullock} J.~S., {Kravtsov} A.~V., {Weinberg} D.~H., 2000, \apj, 539, 517

\bibitem[{{Busha} {et~al}\mbox{.}(2010){Busha}, {Alvarez}, {Wechsler}, {Abel},
  \& {Strigari}}]{Busha2010}
{Busha} M.~T., {Alvarez} M.~A., {Wechsler} R.~H., {Abel} T., {Strigari} L.~E.,
  2010, \apj, 710, 408

\bibitem[{{Busha} {et~al}\mbox{.}(2011){Busha}, {Wechsler}, {Behroozi},
  {Gerke}, {Klypin}, \& {Primack}}]{Busha2011}
{Busha} M.~T., {Wechsler} R.~H., {Behroozi} P.~S., {Gerke} B.~F., {Klypin}
  A.~A., {Primack} J.~R., 2011, \apj, 743, 117

\bibitem[{{Cen} {et~al}\mbox{.}(1994){Cen}, {Gott}, {Ostriker}, \&
  {Turner}}]{Cen1994}
{Cen} R., {Gott}, III J.~R., {Ostriker} J.~P., {Turner} E.~L., 1994, \apj, 423,
  1

\bibitem[{{Collins} {et~al}\mbox{.}(2013){Collins}, {Chapman}, {Rich}, {Ibata},
  {Martin}, {Irwin}, {Bate}, {Lewis}, {Pe{\~n}arrubia}, {Arimoto}, {Casey},
  {Ferguson}, {Koch}, {McConnachie}, \& {Tanvir}}]{Collins2013}
{Collins} M.~L.~M. {et~al.}, 2013, \apj, 768, 172

\bibitem[{{Conroy} {et~al}\mbox{.}(2006){Conroy}, {Wechsler}, \&
  {Kravtsov}}]{Conroy2006}
{Conroy} C., {Wechsler} R.~H., {Kravtsov} A.~V., 2006, \apj, 647, 201

\bibitem[{{Davis} {et~al}\mbox{.}(1985){Davis}, {Efstathiou}, {Frenk}, \&
  {White}}]{Davis1985}
{Davis} M., {Efstathiou} G., {Frenk} C.~S., {White} S.~D.~M., 1985, \apj, 292,
  371

\bibitem[{{de Blok}(2010)}]{deBlok2010}
{de Blok} W.~J.~G., 2010, Advances in Astronomy, 2010

\bibitem[{{Del Popolo}(2012)}]{DelPopolo2012}
{Del Popolo} A., 2012, \mnras, 419, 971

\bibitem[{{Del Popolo} {et~al}\mbox{.}(2014){Del Popolo}, {Lima}, {Fabris}, \&
  {Rodrigues}}]{DelPopolo2014}
{Del Popolo} A., {Lima} J.~A.~S., {Fabris} J.~C., {Rodrigues} D.~C., 2014,
  {arXiv: 1404.3674 [astro-ph]}

\bibitem[{{Di Cintio} {et~al}\mbox{.}(2013){Di Cintio}, {Knebe}, {Libeskind},
  {Brook}, {Yepes}, {Gottl{\"o}ber}, \& {Hoffman}}]{diCintio2013}
{Di Cintio} A., {Knebe} A., {Libeskind} N.~I., {Brook} C., {Yepes} G.,
  {Gottl{\"o}ber} S., {Hoffman} Y., 2013, \mnras, 431, 1220

\bibitem[{{Di Cintio} {et~al}\mbox{.}(2011){Di Cintio}, {Knebe}, {Libeskind},
  {Yepes}, {Gottl{\"o}ber}, \& {Hoffman}}]{diCintio2011}
{Di Cintio} A., {Knebe} A., {Libeskind} N.~I., {Yepes} G., {Gottl{\"o}ber} S.,
  {Hoffman} Y., 2011, \mnras, 417, L74

\bibitem[{{Diemand} {et~al}\mbox{.}(2007){Diemand}, {Kuhlen}, \&
  {Madau}}]{Diemand2007}
{Diemand} J., {Kuhlen} M., {Madau} P., 2007, \apj, 667, 859

\bibitem[{{Efstathiou}(1992)}]{Efstathiou1992}
{Efstathiou} G., 1992, \mnras, 256, 43P

\bibitem[{{Einasto}(1965)}]{Einasto}
{Einasto} J., 1965, Trudy Astrofizicheskogo Instituta Alma-Ata, 5, 87

\bibitem[{{Elbert} {et~al}\mbox{.}(in prep){Elbert}, {Bullock}, \&
  {Kaplinghat}}]{Elbert2014}
{Elbert} O., {Bullock} J.~S., {Kaplinghat} M., in prep, \mnras

\bibitem[{{Flores} \& {Primack}(1994)}]{Flores1994}
{Flores} R.~A., {Primack} J.~R., 1994, \apjl, 427, L1

\bibitem[{{Fraternali} {et~al}\mbox{.}(2009){Fraternali}, {Tolstoy}, {Irwin},
  \& {Cole}}]{Fraternali2009}
{Fraternali} F., {Tolstoy} E., {Irwin} M.~J., {Cole} A.~A., 2009, \aap, 499,
  121

\bibitem[{{Frenk} {et~al}\mbox{.}(1988){Frenk}, {White}, {Davis}, \&
  {Efstathiou}}]{Frenk1988}
{Frenk} C.~S., {White} S.~D.~M., {Davis} M., {Efstathiou} G., 1988, \apj, 327,
  507

\bibitem[{{Gao} {et~al}\mbox{.}(2008){Gao}, {Navarro}, {Cole}, {Frenk},
  {White}, {Springel}, {Jenkins}, \& {Neto}}]{Gao2008}
{Gao} L., {Navarro} J.~F., {Cole} S., {Frenk} C.~S., {White} S.~D.~M.,
  {Springel} V., {Jenkins} A., {Neto} A.~F., 2008, \mnras, 387, 536

\bibitem[{{Garrison-Kimmel} {et~al}\mbox{.}(2014){Garrison-Kimmel},
  {Boylan-Kolchin}, {Bullock}, \& {Lee}}]{ELVIS}
{Garrison-Kimmel} S., {Boylan-Kolchin} M., {Bullock} J.~S., {Lee} K., 2014,
  \mnras, 438, 2578

\bibitem[{{Garrison-Kimmel} {et~al}\mbox{.}(2013){Garrison-Kimmel}, {Rocha},
  {Boylan-Kolchin}, {Bullock}, \& {Lally}}]{Garrison-Kimmel2013}
{Garrison-Kimmel} S., {Rocha} M., {Boylan-Kolchin} M., {Bullock} J.~S., {Lally}
  J., 2013, \mnras, 433, 3539

\bibitem[{{Gelb} \& {Bertschinger}(1994)}]{Gelb1994}
{Gelb} J.~M., {Bertschinger} E., 1994, \apj, 436, 467

\bibitem[{{Giovanelli} {et~al}\mbox{.}(2013){Giovanelli}, {Haynes}, {Adams},
  {Cannon}, {Rhode}, {Salzer}, {Skillman}, {Bernstein-Cooper}, \&
  {McQuinn}}]{Giovanelli2013}
{Giovanelli} R. {et~al.}, 2013, \aj, 146, 15

\bibitem[{{Gnedin}(2000)}]{Gnedin2000}
{Gnedin} N.~Y., 2000, \apj, 542, 535

\bibitem[{{Gottloeber} {et~al}\mbox{.}(2010){Gottloeber}, {Hoffman}, \&
  {Yepes}}]{Gottloeber2010}
{Gottloeber} S., {Hoffman} Y., {Yepes} G., 2010, {arXiv:1005.2687 [astro-ph]}

\bibitem[{{Grcevich} \& {Putman}(2009)}]{Grcevich2009}
{Grcevich} J., {Putman} M.~E., 2009, \apj, 696, 385

\bibitem[{{Gritschneder} \& {Lin}(2013)}]{Gritschneder2013}
{Gritschneder} M., {Lin} D.~N.~C., 2013, \apj, 765, 38

\bibitem[{{Gross} {et~al}\mbox{.}(1998){Gross}, {Somerville}, {Primack},
  {Holtzman}, \& {Klypin}}]{Gross1998}
{Gross} M.~A.~K., {Somerville} R.~S., {Primack} J.~R., {Holtzman} J., {Klypin}
  A., 1998, \mnras, 301, 81

\bibitem[{{Hernquist} {et~al}\mbox{.}(1996){Hernquist}, {Katz}, {Weinberg}, \&
  {Miralda-Escud{\'e}}}]{Hernquist1996}
{Hernquist} L., {Katz} N., {Weinberg} D.~H., {Miralda-Escud{\'e}} J., 1996,
  \apjl, 457, L51

\bibitem[{{Holmberg}(1958)}]{Holmberg1958}
{Holmberg} E., 1958, Meddelanden fran Lunds Astronomiska Observatorium Serie
  II, 136, 1

\bibitem[{{Jardel} \& {Gebhardt}(2012)}]{Jardel2012}
{Jardel} J.~R., {Gebhardt} K., 2012, \apj, 746, 89

\bibitem[{{Jenkins} {et~al}\mbox{.}(2001){Jenkins}, {Frenk}, {White},
  {Colberg}, {Cole}, {Evrard}, {Couchman}, \& {Yoshida}}]{Jenkins2001}
{Jenkins} A., {Frenk} C.~S., {White} S.~D.~M., {Colberg} J.~M., {Cole} S.,
  {Evrard} A.~E., {Couchman} H.~M.~P., {Yoshida} N., 2001, \mnras, 321, 372

\bibitem[{{Katz} \& {White}(1993)}]{Katz1993}
{Katz} N., {White} S.~D.~M., 1993, \apj, 412, 455

\bibitem[{{Kauffmann} {et~al}\mbox{.}(1993){Kauffmann}, {White}, \&
  {Guiderdoni}}]{Kauffmann1993}
{Kauffmann} G., {White} S.~D.~M., {Guiderdoni} B., 1993, \mnras, 264, 201

\bibitem[{{Kazantzidis} {et~al}\mbox{.}(2004){Kazantzidis}, {Mayer},
  {Mastropietro}, {Diemand}, {Stadel}, \& {Moore}}]{Kazantzidis04}
{Kazantzidis} S., {Mayer} L., {Mastropietro} C., {Diemand} J., {Stadel} J.,
  {Moore} B., 2004, \apj, 608, 663

\bibitem[{{Kirby} {et~al}\mbox{.}(2014){Kirby}, {Bullock}, {Boylan-Kolchin},
  {Kaplinghat}, \& {Cohen}}]{Kirby2013}
{Kirby} E.~N., {Bullock} J.~S., {Boylan-Kolchin} M., {Kaplinghat} M., {Cohen}
  J.~G., 2014, {arXiv:1401.1208 [astro-ph]}

\bibitem[{{Klypin} {et~al}\mbox{.}(1999){Klypin}, {Kravtsov}, {Valenzuela}, \&
  {Prada}}]{Klypin1999}
{Klypin} A., {Kravtsov} A.~V., {Valenzuela} O., {Prada} F., 1999, \apj, 522, 82

\bibitem[{{Klypin} {et~al}\mbox{.}(2011){Klypin}, {Trujillo-Gomez}, \&
  {Primack}}]{Klypin2011}
{Klypin} A.~A., {Trujillo-Gomez} S., {Primack} J., 2011, \apj, 740, 102

\bibitem[{{Knollmann} \& {Knebe}(2009)}]{AHF}
{Knollmann} S.~R., {Knebe} A., 2009, \apjs, 182, 608

\bibitem[{{Koch} {et~al}\mbox{.}(2007){Koch}, {Kleyna}, {Wilkinson}, {Grebel},
  {Gilmore}, {Evans}, {Wyse}, \& {Harbeck}}]{Koch2007}
{Koch} A., {Kleyna} J.~T., {Wilkinson} M.~I., {Grebel} E.~K., {Gilmore} G.~F.,
  {Evans} N.~W., {Wyse} R.~F.~G., {Harbeck} D.~R., 2007, \aj, 134, 566

\bibitem[{{Koposov} {et~al}\mbox{.}(2009){Koposov}, {Yoo}, {Rix}, {Weinberg},
  {Macci{\`o}}, \& {Escud{\'e}}}]{Koposov2009}
{Koposov} S.~E., {Yoo} J., {Rix} H.-W., {Weinberg} D.~H., {Macci{\`o}} A.~V.,
  {Escud{\'e}} J.~M., 2009, \apj, 696, 2179

\bibitem[{{Kravtsov}(2010)}]{Kravtsov2010}
{Kravtsov} A., 2010, Advances in Astronomy, 2010

\bibitem[{{Kravtsov} {et~al}\mbox{.}(2004){Kravtsov}, {Berlind}, {Wechsler},
  {Klypin}, {Gottl{\"o}ber}, {Allgood}, \& {Primack}}]{Kravtsov2004}
{Kravtsov} A.~V., {Berlind} A.~A., {Wechsler} R.~H., {Klypin} A.~A.,
  {Gottl{\"o}ber} S., {Allgood} B., {Primack} J.~R., 2004, \apj, 609, 35

\bibitem[{{Kuzio de Naray} \& {Kaufmann}(2011)}]{KuziodeNaray2011}
{Kuzio de Naray} R., {Kaufmann} T., 2011, \mnras, 414, 3617

\bibitem[{{Kuzio de Naray} {et~al}\mbox{.}(2008){Kuzio de Naray}, {McGaugh}, \&
  {de Blok}}]{KuziodeNaray2008}
{Kuzio de Naray} R., {McGaugh} S.~S., {de Blok} W.~J.~G., 2008, \apj, 676, 920

\bibitem[{{Larson} {et~al}\mbox{.}(2011){Larson}, {Dunkley}, {Hinshaw},
  {Komatsu}, {Nolta}, {Bennett}, {Gold}, {Halpern}, {Hill}, {Jarosik}, {Kogut},
  {Limon}, {Meyer}, {Odegard}, {Page}, {Smith}, {Spergel}, {Tucker}, {Weiland},
  {Wollack}, \& {Wright}}]{Larson2011}
{Larson} D. {et~al.}, 2011, \apjs, 192, 16

\bibitem[{{Leaman} {et~al}\mbox{.}(2012){Leaman}, {Venn}, {Brooks},
  {Battaglia}, {Cole}, {Ibata}, {Irwin}, {McConnachie}, {Mendel}, \&
  {Tolstoy}}]{Leaman2012}
{Leaman} R. {et~al.}, 2012, \apj, 750, 33

\bibitem[{{Letarte} {et~al}\mbox{.}(2009){Letarte}, {Chapman}, {Collins},
  {Ibata}, {Irwin}, {Ferguson}, {Lewis}, {Martin}, {McConnachie}, \&
  {Tanvir}}]{Letarte2009}
{Letarte} B. {et~al.}, 2009, \mnras, 400, 1472

\bibitem[{{Lovell} {et~al}\mbox{.}(2013){Lovell}, {Frenk}, {Eke}, {Jenkins},
  {Gao}, \& {Theuns}}]{Lovell2013}
{Lovell} M.~R., {Frenk} C.~S., {Eke} V.~R., {Jenkins} A., {Gao} L., {Theuns}
  T., 2013, {arXiv:1308.1399 [astro-ph]}

\bibitem[{{Lunnan} {et~al}\mbox{.}(2012){Lunnan}, {Vogelsberger}, {Frebel},
  {Hernquist}, {Lidz}, \& {Boylan-Kolchin}}]{Lunnan2012}
{Lunnan} R., {Vogelsberger} M., {Frebel} A., {Hernquist} L., {Lidz} A.,
  {Boylan-Kolchin} M., 2012, \apj, 746, 109

\bibitem[{{Martinez}(2013)}]{Martinez2013}
{Martinez} G.~D., 2013, {arXiv:1309.2641 [astro-ph]}

\bibitem[{{Mateo} {et~al}\mbox{.}(2008){Mateo}, {Olszewski}, \&
  {Walker}}]{Mateo2008}
{Mateo} M., {Olszewski} E.~W., {Walker} M.~G., 2008, \apj, 675, 201

\bibitem[{{McConnachie}(2012)}]{McConnachie2012}
{McConnachie} A.~W., 2012, \aj, 144, 4

\bibitem[{{Moore}(1994)}]{Moore1994}
{Moore} B., 1994, \nat, 370, 629

\bibitem[{{Moore} {et~al}\mbox{.}(1999){Moore}, {Ghigna}, {Governato}, {Lake},
  {Quinn}, {Stadel}, \& {Tozzi}}]{Moore1999}
{Moore} B., {Ghigna} S., {Governato} F., {Lake} G., {Quinn} T., {Stadel} J.,
  {Tozzi} P., 1999, \apjl, 524, L19

\bibitem[{{Moster} {et~al}\mbox{.}(2013){Moster}, {Naab}, \&
  {White}}]{Moster2013}
{Moster} B.~P., {Naab} T., {White} S.~D.~M., 2013, \mnras, 428, 3121

\bibitem[{{Mu{\~n}oz} {et~al}\mbox{.}(2005){Mu{\~n}oz}, {Frinchaboy},
  {Majewski}, {Kuhn}, {Chou}, {Palma}, {Sohn}, {Patterson}, \&
  {Siegel}}]{Munoz2005}
{Mu{\~n}oz} R.~R. {et~al.}, 2005, \apjl, 631, L137

\bibitem[{{Navarro} {et~al}\mbox{.}(1996){Navarro}, {Frenk}, \& {White}}]{NFW}
{Navarro} J.~F., {Frenk} C.~S., {White} S.~D.~M., 1996, \apj, 462, 563

\bibitem[{{Navarro} {et~al}\mbox{.}(2010){Navarro}, {Ludlow}, {Springel},
  {Wang}, {Vogelsberger}, {White}, {Jenkins}, {Frenk}, \&
  {Helmi}}]{Navarro2010}
{Navarro} J.~F. {et~al.}, 2010, \mnras, 402, 21

\bibitem[{{O{\~n}orbe} {et~al}\mbox{.}(2014){O{\~n}orbe}, {Garrison-Kimmel},
  {Maller}, {Bullock}, {Rocha}, \& {Hahn}}]{Onorbe2013}
{O{\~n}orbe} J., {Garrison-Kimmel} S., {Maller} A.~H., {Bullock} J.~S., {Rocha}
  M., {Hahn} O., 2014, \mnras, 437, 1894

\bibitem[{{Okamoto} {et~al}\mbox{.}(2008){Okamoto}, {Gao}, \&
  {Theuns}}]{Okamoto2008}
{Okamoto} T., {Gao} L., {Theuns} T., 2008, \mnras, 390, 920

\bibitem[{{Pe{\~n}arrubia} {et~al}\mbox{.}(2012){Pe{\~n}arrubia}, {Pontzen},
  {Walker}, \& {Koposov}}]{Penarrubia2012}
{Pe{\~n}arrubia} J., {Pontzen} A., {Walker} M.~G., {Koposov} S.~E., 2012,
  \apjl, 759, L42

\bibitem[{{Planck Collaboration} {et~al}\mbox{.}(2013){Planck Collaboration},
  {Ade}, {Aghanim}, {Armitage-Caplan}, {Arnaud}, {Ashdown}, {Atrio-Barandela},
  {Aumont}, {Baccigalupi}, {Banday}, \& et~al.}]{PlanckCosmo}
{Planck Collaboration} {et~al.}, 2013, {arXiv:1303.5076 [astro-ph]}

\bibitem[{{Polisensky} \& {Ricotti}(2014)}]{Polisensky2013}
{Polisensky} E., {Ricotti} M., 2014, \mnras, 437, 2922

\bibitem[{{Pontzen} \& {Governato}(2012)}]{Pontzen2012}
{Pontzen} A., {Governato} F., 2012, \mnras, 421, 3464

\bibitem[{{Purcell} \& {Zentner}(2012)}]{Purcell2012}
{Purcell} C.~W., {Zentner} A.~R., 2012, \jcap, 12, 7

\bibitem[{{Rhode} {et~al}\mbox{.}(2013){Rhode}, {Salzer}, {Haurberg}, {Van
  Sistine}, {Young}, {Haynes}, {Giovanelli}, {Cannon}, {Skillman}, {McQuinn},
  \& {Adams}}]{Rhode2013}
{Rhode} K.~L. {et~al.}, 2013, \aj, 145, 149

\bibitem[{{Rocha} {et~al}\mbox{.}(2013){Rocha}, {Peter}, {Bullock},
  {Kaplinghat}, {Garrison-Kimmel}, {O{\~n}orbe}, \& {Moustakas}}]{Rocha2013}
{Rocha} M., {Peter} A.~H.~G., {Bullock} J.~S., {Kaplinghat} M.,
  {Garrison-Kimmel} S., {O{\~n}orbe} J., {Moustakas} L.~A., 2013, \mnras, 430,
  81

\bibitem[{{Rodr{\'{\i}}guez-Puebla}
  {et~al}\mbox{.}(2013){Rodr{\'{\i}}guez-Puebla}, {Avila-Reese}, \&
  {Drory}}]{Rodriguez2013}
{Rodr{\'{\i}}guez-Puebla} A., {Avila-Reese} V., {Drory} N., 2013, \apj, 773,
  172

\bibitem[{{Sawala} {et~al}\mbox{.}(2014){Sawala}, {Frenk}, {Fattahi},
  {Navarro}, {Bower}, {Crain}, {Dalla Vecchia}, {Furlong}, {Jenkins},
  {McCarthy}, {Qu}, {Schaller}, {Schaye}, \& {Theuns}}]{Sawala2014}
{Sawala} T. {et~al.}, 2014, {arXiv:1404.3724 [astro-ph]}

\bibitem[{{Simon} \& {Geha}(2007)}]{Simon2007}
{Simon} J.~D., {Geha} M., 2007, \apj, 670, 313

\bibitem[{{Slater} {et~al}\mbox{.}(2011){Slater}, {Bell}, \&
  {Martin}}]{Slater2011}
{Slater} C.~T., {Bell} E.~F., {Martin} N.~F., 2011, \apjl, 742, L14

\bibitem[{{Somerville}(2002)}]{Somerville2002}
{Somerville} R.~S., 2002, \apjl, 572, L23

\bibitem[{{Springel} {et~al}\mbox{.}(2008){Springel}, {Wang}, {Vogelsberger},
  {Ludlow}, {Jenkins}, {Helmi}, {Navarro}, {Frenk}, \& {White}}]{Aquarius}
{Springel} V. {et~al.}, 2008, \mnras, 391, 1685

\bibitem[{{Springel} {et~al}\mbox{.}(2005){Springel}, {White}, {Jenkins},
  {Frenk}, {Yoshida}, {Gao}, {Navarro}, {Thacker}, {Croton}, {Helly},
  {Peacock}, {Cole}, {Thomas}, {Couchman}, {Evrard}, {Colberg}, \&
  {Pearce}}]{Springel2005}
{Springel} V. {et~al.}, 2005, \nat, 435, 629

\bibitem[{{Stanimirovi{\'c}} {et~al}\mbox{.}(2004){Stanimirovi{\'c}},
  {Staveley-Smith}, \& {Jones}}]{Stanimirovic2004}
{Stanimirovi{\'c}} S., {Staveley-Smith} L., {Jones} P.~A., 2004, \apj, 604, 176

\bibitem[{{Strigari} {et~al}\mbox{.}(2008){Strigari}, {Bullock}, {Kaplinghat},
  {Simon}, {Geha}, {Willman}, \& {Walker}}]{Strigari2008}
{Strigari} L.~E., {Bullock} J.~S., {Kaplinghat} M., {Simon} J.~D., {Geha} M.,
  {Willman} B., {Walker} M.~G., 2008, \nat, 454, 1096

\bibitem[{{Teyssier} {et~al}\mbox{.}(2012){Teyssier}, {Johnston}, \&
  {Kuhlen}}]{Teyssier2012}
{Teyssier} M., {Johnston} K.~V., {Kuhlen} M., 2012, \mnras, 426, 1808

\bibitem[{{Thoul} \& {Weinberg}(1996)}]{Thoul1996}
{Thoul} A.~A., {Weinberg} D.~H., 1996, \apj, 465, 608

\bibitem[{{Tollerud} {et~al}\mbox{.}(2012){Tollerud}, {Beaton}, {Geha},
  {Bullock}, {Guhathakurta}, {Kalirai}, {Majewski}, {Kirby}, {Gilbert},
  {Yniguez}, {Patterson}, {Ostheimer}, {Cooke}, {Dorman}, {Choudhury}, \&
  {Cooper}}]{Tollerud2012}
{Tollerud} E.~J. {et~al.}, 2012, \apj, 752, 45

\bibitem[{{Tollerud} {et~al}\mbox{.}(2011){Tollerud}, {Boylan-Kolchin},
  {Barton}, {Bullock}, \& {Trinh}}]{Tollerud2011}
{Tollerud} E.~J., {Boylan-Kolchin} M., {Barton} E.~J., {Bullock} J.~S., {Trinh}
  C.~Q., 2011, \apj, 738, 102

\bibitem[{{Tollerud} {et~al}\mbox{.}(2014){Tollerud}, {Boylan-Kolchin}, \&
  {Bullock}}]{Tollerud2014}
{Tollerud} E.~J., {Boylan-Kolchin} M., {Bullock} J.~S., 2014, {arXiv:1403.6469
  [astro-ph]}

\bibitem[{{Trachternach} {et~al}\mbox{.}(2008){Trachternach}, {de Blok},
  {Walter}, {Brinks}, \& {Kennicutt}}]{Trachternach2008}
{Trachternach} C., {de Blok} W.~J.~G., {Walter} F., {Brinks} E., {Kennicutt},
  Jr. R.~C., 2008, \aj, 136, 2720

\bibitem[{{Vale} \& {Ostriker}(2004)}]{Vale2004}
{Vale} A., {Ostriker} J.~P., 2004, \mnras, 353, 189

\bibitem[{{van der Marel} {et~al}\mbox{.}(2012){van der Marel}, {Fardal},
  {Besla}, {Beaton}, {Sohn}, {Anderson}, {Brown}, \&
  {Guhathakurta}}]{Marel2012}
{van der Marel} R.~P., {Fardal} M., {Besla} G., {Beaton} R.~L., {Sohn} S.~T.,
  {Anderson} J., {Brown} T., {Guhathakurta} P., 2012, \apj, 753, 8

\bibitem[{{Vera-Ciro} {et~al}\mbox{.}(2013){Vera-Ciro}, {Helmi}, {Starkenburg},
  \& {Breddels}}]{Vera-Ciro2013}
{Vera-Ciro} C.~A., {Helmi} A., {Starkenburg} E., {Breddels} M.~A., 2013,
  \mnras, 428, 1696

\bibitem[{{Vogelsberger} {et~al}\mbox{.}(2012){Vogelsberger}, {Zavala}, \&
  {Loeb}}]{Vogelsberger2012}
{Vogelsberger} M., {Zavala} J., {Loeb} A., 2012, \mnras, 423, 3740

\bibitem[{{Walker} {et~al}\mbox{.}(2009){Walker}, {Mateo}, \&
  {Olszewski}}]{Walker2009}
{Walker} M.~G., {Mateo} M., {Olszewski} E.~W., 2009, \aj, 137, 3100

\bibitem[{{Walker} \& {Pe{\~n}arrubia}(2011)}]{Walker2011}
{Walker} M.~G., {Pe{\~n}arrubia} J., 2011, \apj, 742, 20

\bibitem[{{Wambsganss} {et~al}\mbox{.}(2004){Wambsganss}, {Bode}, \&
  {Ostriker}}]{Wambsganss2004}
{Wambsganss} J., {Bode} P., {Ostriker} J.~P., 2004, \apjl, 606, L93

\bibitem[{{Wang} {et~al}\mbox{.}(2012){Wang}, {Frenk}, {Navarro}, {Gao}, \&
  {Sawala}}]{Wang2012}
{Wang} J., {Frenk} C.~S., {Navarro} J.~F., {Gao} L., {Sawala} T., 2012, \mnras,
  424, 2715

\bibitem[{{Warren} {et~al}\mbox{.}(1992){Warren}, {Quinn}, {Salmon}, \&
  {Zurek}}]{Warren1992}
{Warren} M.~S., {Quinn} P.~J., {Salmon} J.~K., {Zurek} W.~H., 1992, \apj, 399,
  405

\bibitem[{{Weiner} {et~al}\mbox{.}(2006){Weiner}, {Willmer}, {Faber},
  {Melbourne}, {Kassin}, {Phillips}, {Harker}, {Metevier}, {Vogt}, \&
  {Koo}}]{Weiner2006}
{Weiner} B.~J. {et~al.}, 2006, \apj, 653, 1027

\bibitem[{{Willman} {et~al}\mbox{.}(2005){Willman}, {Blanton}, {West},
  {Dalcanton}, {Hogg}, {Schneider}, {Wherry}, {Yanny}, \&
  {Brinkmann}}]{Willman2005}
{Willman} B. {et~al.}, 2005, \aj, 129, 2692

\bibitem[{{Wolf} {et~al}\mbox{.}(2010){Wolf}, {Martinez}, {Bullock},
  {Kaplinghat}, {Geha}, {Mu{\~n}oz}, {Simon}, \& {Avedo}}]{Wolf2010}
{Wolf} J., {Martinez} G.~D., {Bullock} J.~S., {Kaplinghat} M., {Geha} M.,
  {Mu{\~n}oz} R.~R., {Simon} J.~D., {Avedo} F.~F., 2010, \mnras, 406, 1220

\bibitem[{{Woo} {et~al}\mbox{.}(2008){Woo}, {Courteau}, \& {Dekel}}]{Woo2008}
{Woo} J., {Courteau} S., {Dekel} A., 2008, \mnras, 390, 1453

\bibitem[{{Yniguez} {et~al}\mbox{.}(2013){Yniguez}, {Garrison-Kimmel},
  {Boylan-Kolchin}, \& {Bullock}}]{Yniguez2013}
{Yniguez} B., {Garrison-Kimmel} S., {Boylan-Kolchin} M., {Bullock} J.~S., 2013,
  {arXiv:1305.0560 [astro-ph]}

\bibitem[{{Zaggia} {et~al}\mbox{.}(2011){Zaggia}, {Held}, {Sommariva},
  {Momany}, {Saviane}, \& {Rizzi}}]{Zaggia2011}
{Zaggia} S., {Held} E.~V., {Sommariva} V., {Momany} Y., {Saviane} I., {Rizzi}
  L., 2011, in EAS Publications Series, Vol.~48, EAS Publications Series,
  {Koleva} M., {Prugniel} P., {Vauglin} I., eds., pp. 215--216

\bibitem[{{Zavala} {et~al}\mbox{.}(2013){Zavala}, {Vogelsberger}, \&
  {Walker}}]{Zavala2013}
{Zavala} J., {Vogelsberger} M., {Walker} M.~G., 2013, \mnras, 431, L20

\bibitem[{{Zentner} \& {Bullock}(2003)}]{Zentner2003}
{Zentner} A.~R., {Bullock} J.~S., 2003, \apj, 598, 49

\bibitem[{{Zolotov} {et~al}\mbox{.}(2012){Zolotov}, {Brooks}, {Willman},
  {Governato}, {Pontzen}, {Christensen}, {Dekel}, {Quinn}, {Shen}, \&
  {Wadsley}}]{Zolotov2012}
{Zolotov} A. {et~al.}, 2012, \apj, 761, 71

\end{thebibliography}

\section*{Appendix A:  Numerical Convergence}
\label{sec:restest}
\begin{figure}
\centering
\includegraphics[width=0.49\textwidth]{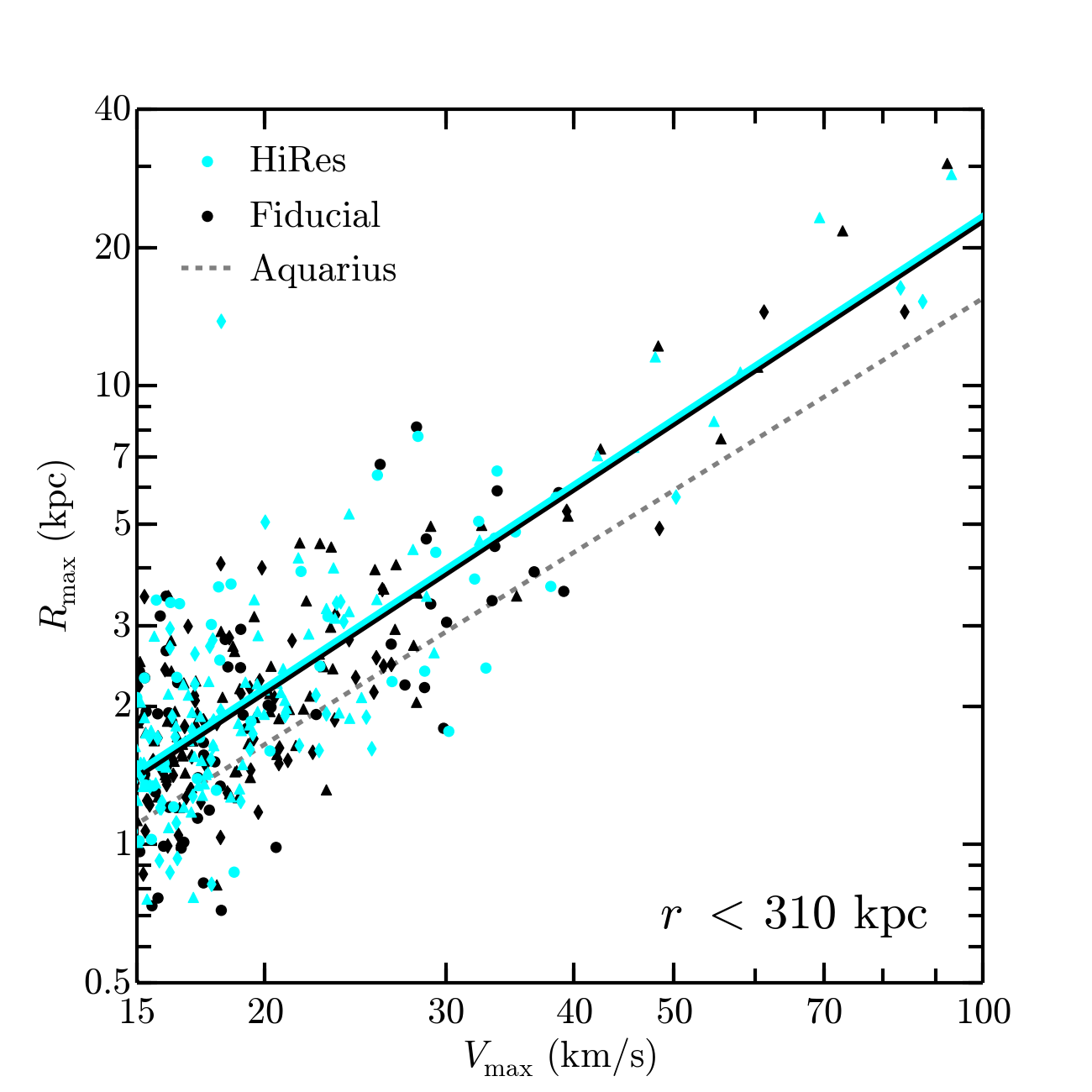}
\caption{Resolution test comparing subhalos within 310~kpc of three of the
isolated analogs in the ELVIS Suite, iKauket (circles), iHall (triangles), and iScylla (diamonds),
at the standard resolution of the ELVIS suite (black points) and with eight times better 
mass resolution (cyan points); the fits to the data, weighted by $\vmax$, are also plotted.  
The normalizations of the fits to halos with $\vmax > 15~\kms$ and $\rmax > 0.5~\kpc$ 
agree to within $3\%$, indicating that our results are not affected by numerical errors.  The 
dashed grey line plots the relation found in \citet{Aquarius};  the offset 
($\sim20\%$) is consistent with the updated $\sigma_8$ used in the 
ELVIS cosmology.  We also find nearly identical relations using halo catalogs 
produced by \texttt{AHF}.}
\label{fig:restest}
\end{figure}

Three of the isolated hosts in the ELVIS Suite were re-simulated with eight times 
better mass resolution than the fiducial runs ($m_{\rm p} = 2.35\times10^4~\msun$) 
and with a $z=0$ softening length of $70$~pc for the high resolution particles.  
Although the individual halo properties vary slightly between these HiRes 
simulations and the fiducial analogs, as expected from \citet{Onorbe2013}, we 
use those simulations here to determine the limits of our full sample.  In 
Figure~\ref{fig:restest}, we plot the relationship between $\rmax$ and $\vmax$ 
for subhalos within 310~kpc of these three hosts.  We use 310~kpc to include 
a large subhalo that, owing to phase differences between the resolutions, 
is beyond 300~kpc at the standard resolution.  Subhalos from the HiRes simulations 
are shown as cyan points and those from the standard resolution runs are plotted in 
black; the symbol types indicate the three host halos.

Fits to both of these populations, including only halos with 
$\vmax > 15~\kms$ and $\rmax > 0.5~\kpc$, are also plotted in Figure~\ref{fig:restest}.  
The power law given by Equation~\ref{eqn:310fit} fits both populations well, 
with a difference in the normalizations of less than 3\%, indicating that our results 
are robust to resolution errors.  We have also checked that our results do not depend 
on the specific halo finder by repeating this analysis with halo catalogs produced by 
\texttt{Amiga Halo Finder} \citep{AHF}, which locates spherical overdensities in the 
three-dimensional matter distribution~--~the normalizations differ by $5\%$ at most.   
\texttt{Rockstar} also appears to misidentify $\rmax$ for a single small halo in the 
high resolution run; this halo, however, is not used in the full analysis and 
does not strongly bias the fit. 

\section*{Appendix B:  Density Profiles}
\label{sec:profiles}
Rather than individually fit profiles to each subhalo (an inaccurate approach, 
due to the insufficient resolution at low radii and relatively small differences 
in the profiles near $\rmax$), we perform our analysis using three Einasto 
profiles ($\alpha = 0.15,~0.18$,~and~$0.28$).  As shown in \citet{Aquarius}, 
an Einasto profile with $\alpha$ fixed at $0.18$ is a better fit to most subhalos 
than a standard NFW profile~--~we therefore focus our efforts on this profile.  
Though a comprehensive analysis of the distribution of best-fit shape parameters 
of ultra-high resolution subhalos and field dwarfs does not exist in the literature, 
$\alpha = 0.15~\mathrm{and}~0.28$ are the extreme values plotted in
\citet{Aquarius} and we therefore consider those shape parameters as an estimate
of appropriate scatter.

For a given $\alpha$, the circular velocity may be expressed as a function of 
$\rmax$ and $\vmax$, parameters which are robustly determined for the halos 
considered in this work (see Figure~\ref{fig:restest}).  For the Einasto profile, 
\begin{multline}
\frac{V_{\rm circ}^2(r)}{\vmax^2} = \frac{4\pi/\alpha}{A(\alpha) B(\alpha)}\exp\left(\frac{2-\log(8)+3\log(\alpha)}{\alpha}\right)  \\                 \times \gamma\left(\frac{3}{\alpha},\frac{2}{\alpha}\left(\frac{A(\alpha) r}{\rmax}\right)^\alpha\right)\frac{\rmax}{r},
\label{eqn:einastoVcirc}
\end{multline}
where $\gamma(x,y)$ is the lower incomplete gamma function.  $A(\alpha)$ and 
$B(\alpha)$ relate $\vmax$ and $\rmax$ to $r_{-2}$ and $\rho_{-2}$, the radius 
at which the log slope of the density profile is $-2$ and the density at that radius, via
\begin{equation}
\begin{aligned}
\rmax &= A(\alpha) r_{-2} \\
\vmax^2 &= B(\alpha) G \rho_{-2} r_{-2}^2,
\end{aligned}
\label{eqn:EinastoVc}
\end{equation}

By finding the maximum of Equation~\ref{eqn:einastoVcirc}, one can show that 
$A(\alpha)$ is given by the root of
\begin{equation}
e^{-2x^\alpha/\alpha}\alpha^{\frac{\alpha-3}{\alpha}}x^3 - 8^{-1/\alpha}\gamma\left(\frac{3}{\alpha},\frac{2x^\alpha}{\alpha}\right) = 0,
\end{equation}
where $x = r/r_{-2}$.  $B(\alpha)$ may then be obtained by directly calculating 
$V_{\rm circ}(r)$ at $\rmax$.  For $0 < \alpha < 1$, $A(\alpha)$ and $B(\alpha)$ 
are well fit by two-power functions:
\begin{equation}
\begin{aligned}
A(\alpha) &= 1.715\alpha^{-0.00183}(\alpha + 0.0817)^{-0.179488} \\
B(\alpha) &= 9.529\alpha^{-0.00635}(\alpha + 0.3036)^{-0.206886}
\end{aligned}
\label{eqn:EinastoABfits}
\end{equation}

\begin{figure}
\centering
\includegraphics[width=0.49\textwidth]{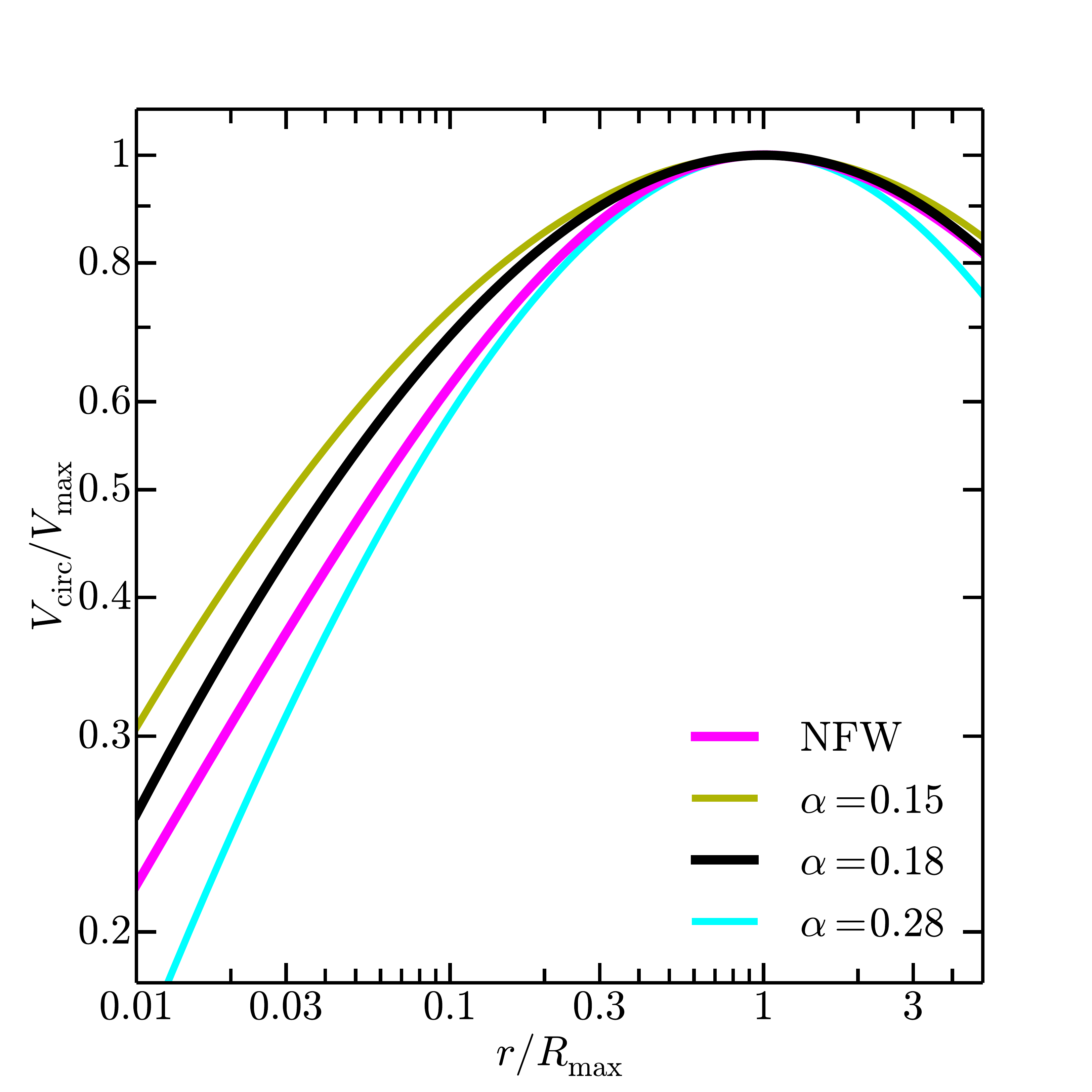}
\caption{Circular velocities profiles, normalized by $\rmax$ and $\vmax$ for
the three shape parameters considered above:  $\alpha = 0.15$ (dark yellow), 
$\alpha = 0.18$ (black), and $\alpha=0.28$ (cyan), along with that of an NFW 
profile (magenta).  Smaller shape parameters result in denser halos, and therefore 
more massive failures.}
\label{fig:vcirc_comp}
\end{figure}

In Figure~\ref{fig:vcirc_comp}, we compare the resultant circular velocity curves 
for these three shape parameters, along with that of an NFW profile.  Smaller
values of $\alpha$ result in more mass near the center of halos and therefore 
lead to more unaccounted-for objects and massive failures in the simulations.

\label{lastpage}
\end{document}